\documentclass[3p]{elsarticle}

\usepackage{amsmath}
\usepackage{amsthm}
\usepackage{amssymb}
\usepackage{graphicx}
\usepackage{esint}
\usepackage{xcolor}

\allowdisplaybreaks

\usepackage{lineno}
\PassOptionsToPackage{hyphens}{url}
\usepackage[pdfencoding=auto,psdextra,pdfauthor=author]{hyperref}
\usepackage{bookmark}

\usepackage{chngcntr}

\biboptions{numbers,sort&compress}

\modulolinenumbers[1]

\journal{Physics Letters A}

\begin{document}

\begin{frontmatter}

\title{Quantum tunneling of three-spine solitons through excentric barriers}

\author[address1]{Danko D. Georgiev\corref{mycorrespondingauthor}}
\ead{danko.georgiev@mail.bg}
\cortext[mycorrespondingauthor]{Corresponding author}

\author[address2]{James F. Glazebrook}
\ead{jfglazebrook@eiu.edu}

\address[address1]{Institute for Advanced Study, 30 Vasilaki Papadopulu Str., Varna 9010, Bulgaria}
\address[address2]{Department of Mathematics, Eastern Illinois University, 600 Lincoln Ave., Charleston, Illinois 61920-3099, USA}

\begin{abstract}
Macromolecular protein complexes catalyze essential physiological processes that sustain life. Various interactions between protein subunits could increase the effective mass of certain peptide groups, thereby compartmentalizing protein $\alpha$-helices. Here, we study the differential effects of applied massive barriers upon the soliton-assisted energy transport within proteins. We demonstrate that excentric barriers, localized onto a single spine in the protein $\alpha$-helix, reflect or trap three-spine solitons as effectively as concentric barriers with comparable total mass. Furthermore, wider protein solitons, whose energy is lower, require heavier massive barriers for soliton reflection or trapping. Regulation of energy transport, delivery and utilization at protein active sites could thus be achieved through control of the soliton width, or of the effective mass of the protein subunits.
\end{abstract}


\date{July 2, 2022}

\end{frontmatter}


\section*{Highlights}

\begin{itemize}

\item Three-spine solitons provide a physical mechanism for transportation of amide~I energy inside protein $\alpha$-helices.

\item Macromolecular protein assembly could increase effective mass thereby compartmentalizing protein $\alpha$-helices.

\item Three-spine solitons are able to tunnel through, reflect from or remain trapped by excentric massive barriers.

\item Wider three-spine solitons have lower energy, move at slower speed and manifest increased tunneling probability.

\item Regulation of total barrier mass by protein subunits could control soliton tunneling, reflection or trapping.

\end{itemize}

\pagebreak

\section{Introduction}

Proteins sustain life through a catalysis of biochemical reactions within
living systems \cite{Rodwell2018,Chiesa2020,Stollar2020}. Different
biological functions are performed by specialized macromolecular protein
complexes, which are assembled by multiple protein subunits \cite{ODonoghue2010,Durand2013,Larochelle2015}.
Important examples of functionally active macromolecular protein complexes
include enzymes, motor proteins, active pumps, biomolecule transporters,
and transmembrane ion channels \cite{Creighton1993,Kessel2018,Georgiev2020c}.
The $\alpha$-helix is a ubiquitous secondary structural element in large proteins \cite{Pauling1951,Sivaramakrishnan2008},
which provides an ordered biological environment for the transport
of metabolic energy in the form of molecular solitons: specifically, a model for molecular quasiparticles, as first proposed
by Alexander Davydov \cite{Davydov1976,Davydov1979,Davydov1979b,Davydov1981,Davydov1982,Davydov1986}.
The biophysical conditions for the creation and propagation of molecular
solitons in protein $\alpha$-helices were further studied both for
the continuum case, as modeled with nonlinear Schr\"{o}dinger equation for
excitons \cite{Brizhik1983,Brizhik1988,Brizhik1993,Kivshar1989},
and for the discrete case modeled by a system of coupled differential
equations obtained from the linear Schr\"{o}dinger equation for the composite
system of excitons and phonons \cite{Brizhik2004,Brizhik2006,Brizhik2019,Georgiev2019a,Georgiev2020a,Georgiev2022}.

The general Hamiltonian for Davydov's model is given as a sum of three
terms describing excitons, phonons and the corresponding exciton--phonon
interaction \cite{Davydov1978,Hyman1981,Davydov1979,Davydov1982,Luo2017,Brizhik2019,Georgiev2022}
\begin{equation}
\hat{H}=\hat{H}_{\textrm{ex}}+\hat{H}_{\textrm{ph}}+\hat{H}_{\textrm{int}}\label{eq:Hamiltonian}
\end{equation}
with
\begin{align}
\hat{H}_{\textrm{ex}} & =\sum_{n,\alpha}\left[E_{0}\hat{a}_{n,\alpha}^{\dagger}\hat{a}_{n,\alpha}-J_{1}\left(\hat{a}_{n,\alpha}^{\dagger}\hat{a}_{n+1,\alpha}+\hat{a}_{n,\alpha}^{\dagger}\hat{a}_{n-1,\alpha}\right)+J_{2}\left(\hat{a}_{n,\alpha}^{\dagger}\hat{a}_{n,\alpha+1}+\hat{a}_{n,\alpha}^{\dagger}\hat{a}_{n,\alpha-1}\right)\right]\\
\hat{H}_{\textrm{ph}} & =\frac{1}{2}\sum_{n,\alpha}\left[\frac{\hat{p}_{n,\alpha}^{2}}{M_{n,\alpha}}+w_{1}\left(\hat{u}_{n+1,\alpha}-\hat{u}_{n,\alpha}\right)^{2}+w_{2}\left(\hat{u}_{n,\alpha+1}-\hat{u}_{n,\alpha}\right)^{2}\right]\\
\hat{H}_{\textrm{int}} & =\sum_{n,\alpha}\left[\chi_{l}\left(\hat{u}_{n,\alpha}-\hat{u}_{n-1,\alpha}\right)+\chi_{r}\left(\hat{u}_{n+1,\alpha}-\hat{u}_{n,\alpha}\right)\right]\hat{a}_{n,\alpha}^{\dagger}\hat{a}_{n,\alpha}
\end{align}
where $E_{0}=0.2$~eV is the energy of the amide~I exciton, $J_{1}=967.4$~$\mu$eV
is the nearest neighbor longitudinal dipole--dipole coupling energy
along a spine of hydrogen-bonded peptide groups \cite{Nevskaya1976},
$J_{2}=1535.4$~$\mu$eV is the nearest neighbor lateral dipole--dipole
coupling energy between spines \cite{Nevskaya1976}, $\hat{a}_{n}^{\dagger}$
and $\hat{a}_{n}$ are bosonic amide~I exciton creation and annihilation
operators, $w_{1}=13$ N/m is the spring constant of the longitudinal
hydrogen bonds in the lattice of peptide groups \cite{Itoh1972},
$w_{2}=2w_{1}$ is the spring constant of the lateral coupling of
the lattice of peptide groups due to covalent bonds in the protein
backbone \cite{Savin1993,Georgiev2022}, $M_{n,\alpha}$ is the mass,
$\hat{p}_{n,\alpha}$~is the momentum operator and $\hat{u}_{n,\alpha}$
is the displacement operator from the equilibrium position of the
lattice site $n,\alpha$; $\chi_{r}$ and $\chi_{l}$ are anharmonic
parameters arising from the coupling between the amide~I exciton
and the phonon lattice displacements, respectively, to the right or
to the left along the spines, $n$ is an index that counts the peptide
groups along the protein spines, and $\alpha\in\left\{ 0,1,2\right\}$
is an index subject to modulo 3 arithmetic denoting each of the three
spines in the protein $\alpha$-helix \cite{Georgiev2022}.
The modular arithmetic for the spine index $\alpha$ arises due to the cyclic arrangement of the three spines with respect to the central axis of the protein $\alpha$-helix, namely,
the spine with index $\alpha=0$ interacts with the spine to the left, $(\alpha-1)~\textrm{mod}~3 = 2$, or the spine to the right, $(\alpha+1)~\textrm{mod}~3 = 1$,
the spine with index $\alpha=1$ interacts with the spine to the left, $(\alpha-1)~\textrm{mod}~3 = 0$, or the spine to the right, $(\alpha+1)~\textrm{mod}~3 = 2$, and
the spine with index $\alpha=2$ interacts with the spine to the left, $(\alpha-1)~\textrm{mod}~3 = 1$, or the spine to the right, $(\alpha+1)~\textrm{mod}~3 = 0$.

Historically, the three-spine model of a protein $\alpha$-helix was first developed in 1978 by Davydov, Eremko and Sergienko \cite{Davydov1978}, and then numerically simulated in 1981 by Hyman, McLaughlin and Scott \cite{Hyman1981}.
Improvement of the model with account of the helicity of $\alpha$-helical proteins was achieved in 2004 by Brizhik, Eremko, Piette and Zakrzewski \cite{Brizhik2004} who have found that there exist several types of solitons of different parameters and energies such that the lowest energy entangled soliton has the energy which is more than an order lower than the energy of the soliton in the three-spine model without helicity.
Here, we elaborate on those previous works by considering the quantum dynamics of multiple amide~I quanta. In order to study the effects of excentric massive barriers in the protein $\alpha$-helix, it is mandatory to use the full three-spine model because excentric placement of the barrier is impossible on a quasi-one-dimensional system with single chain \cite{Scott1992,Georgiev2019b,Georgiev2022}.

The quantum equations of motion for a given number $Q$ of amide~I exciton
quanta are obtainable from the linear Schr\"{o}dinger equation for
the composite system
\begin{equation}
\imath\hbar\,\frac{d}{dt}|\Psi(t)\rangle=\hat{H}\,|\Psi(t)\rangle\label{eq:schrodinger}
\end{equation}
using the adiabatic approximation \cite{Brizhik2003}
\begin{equation}
|\Psi(t)\rangle=|\psi_{\textrm{ex}}(t)\rangle|\psi_{\textrm{ph}}(t)\rangle
\label{eq:ansatz}
\end{equation}
based on the restricted bosonic Hartree--Fock ansatz \cite{Pitaevskii2003} for amide~I excitons introduced by Zolotaryuk for studying multiquantal Davydov solitons \cite{Zolotaryuk1988}
\begin{equation}
|\psi_{\textrm{ex}}(t)\rangle =\frac{1}{\sqrt{Q!}}\left[\sum_{n,\alpha}a_{n,\alpha}(t)\hat{a}_{n,\alpha}^{\dagger}\right]^{Q}|0_{\textrm{ex}}\rangle
\label{eq:Hartree}
\end{equation}
where $|0_{\textrm{ex}}\rangle$~is the vacuum state of amide~I excitons,
and Glauber coherent phonon state of the lattice \cite{Glauber1963a,Glauber1963b,Davydov1979,Scott1992}
\begin{equation}
|\psi_{\textrm{ph}}(t)\rangle  =e^{-\frac{\imath}{\hbar}\sum_{n',\alpha'}\left(b_{n',\alpha'}(t)\hat{p}_{n',\alpha'}-c_{n',\alpha'}(t)\hat{u}_{n',\alpha'}\right)}|0_{\textrm{ph}}\rangle\label{eq:coherent}
\end{equation}
where $|0_{\textrm{ph}}\rangle$~is the vacuum state of lattice phonons.
It is worth noting that the phonon summation indices $n',\alpha'$ are
independent of the exciton summation indices $n,\alpha$
when the tensor product $|\psi_{\textrm{ex}}(t)\rangle|\psi_{\textrm{ph}}(t)\rangle$ is formed.

Normalization of the state $|\Psi(t)\rangle$ is enforced by setting the inner product to unity
\begin{equation}
\langle \Psi(t)|\Psi(t)\rangle = \left(\sum_{n,\alpha}\left|a_{n,\alpha}\right|^{2}\right)^{Q} = 1
\end{equation}
which implies that $\sum_{n,\alpha}\left|a_{n,\alpha}\right|^{2}=1$.

From the ansatz \eqref{eq:ansatz}, we can compute the following expectation values
\begin{align}
Q\,|a_{n,\alpha}|^{2} & =\langle\Psi(t)|\hat{a}_{n,\alpha}^{\dagger}\hat{a}_{n,\alpha}|\Psi(t)\rangle\\
b_{n,\alpha} & =\langle\Psi(t)|\hat{u}_{n,\alpha}|\Psi(t)\rangle\\
c_{n,\alpha} & =\langle\Psi(t)|\hat{p}_{n,\alpha}|\Psi(t)\rangle
\end{align}
The time dynamics for the amide~I exciton quantum probability amplitudes
could be obtained from the Schr\"{o}dinger equation using the inner product
\begin{equation}
\frac{1}{\sqrt{Q!}}\langle\psi_{\textrm{ph}}(t)|\langle0_{\textrm{ex}}|\left(\hat{a}_{n,\alpha}\right)^{Q}\imath\hbar\frac{d}{dt}|\Psi(t)\rangle=\frac{1}{\sqrt{Q!}}\langle\psi_{\textrm{ph}}(t)|\langle0_{\textrm{ex}}|\left(\hat{a}_{n,\alpha}\right)^{Q}\hat{H}|\Psi(t)\rangle
\end{equation}
whereas the time dynamics of the expectation values of the phonon
lattice displacement and momentum operators could be obtained from
the generalized Ehrenfest theorem using the respective commutators
with the total Hamiltonian \eqref{eq:Hamiltonian}
\begin{align}
\imath\hbar\frac{d}{dt}b_{n,\alpha} & =\langle\Psi(t)|\left[\hat{u}_{n,\alpha},\hat{H}\right]|\Psi(t)\rangle\label{eq:Ehrenfest-1}\\
\imath\hbar\frac{d}{dt}c_{n,\alpha} & =\langle\Psi(t)|\left[\hat{p}_{n,\alpha},\hat{H}\right]|\Psi(t)\rangle\label{eq:Ehrenfest-2}
\end{align}
After straightforward but tedious quantum calculations (for detailed derivations
see \cite{Kerr1987,Kerr1990,Georgiev2020a,Georgiev2022}),
one finds that $c_{n,\alpha} = M_{n,\alpha} \frac{d}{dt}b_{n,\alpha}$ and
arrives at the following system of gauge transformed equations
\begin{align}
\imath\hbar\frac{d}{dt}a_{n,\alpha} & =-J_{1}\left(a_{n+1,\alpha}+a_{n-1,\alpha}\right)+J_{2}\left(a_{n,\alpha+1}+a_{n,\alpha-1}\right)+\left[\chi_{l}\left(b_{n,\alpha}-b_{n-1,\alpha}\right)+\chi_{r}\left(b_{n+1,\alpha}-b_{n,\alpha}\right)\right]a_{n,\alpha}
\label{eq:full-1}\\
M_{n,\alpha}\frac{d^{2}}{dt^{2}}b_{n,\alpha} & = w_{1}\left(b_{n-1,\alpha}-2b_{n,\alpha}+b_{n+1,\alpha}\right)+w_{2}\left(b_{n,\alpha-1}-2b_{n,\alpha}+b_{n,\alpha+1}\right)\nonumber \\
 & \qquad +Q\left[\chi_{r}\left(\left|a_{n,\alpha}\right|^{2}-\left|a_{n-1,\alpha}\right|^{2}\right)+\chi_{l}\left(\left|a_{n+1,\alpha}\right|^{2}-\left|a_{n,\alpha}\right|^{2}\right)\right]
\label{eq:full-2}
\end{align}
We adopt the latter system of ordinary differential equations as the
starting point of the present computational study.
The main research question that we address is the differential effects of concentric or excentric massive barriers upon the soliton-assisted energy transport within proteins.

\section{Model parameters and initial conditions}

The dynamics of the solitons were simulated in an 18-nm-long protein $\alpha$-helix
with $n_{\max}=40$ lattice sites. The left and right coupling parameters
for the exciton--phonon interaction were considered to be isotropic,
$\chi_{l}=\chi_{r}=35$ pN \cite{Georgiev2022}.
The energy released by hydrolysis of a single adenosine triphosphate (ATP) molecule is experimentally measured to exceed $0.6$ eV \cite{Barclay2020,Kammermeier1982,Weiss2005}, which is sufficient to excite 3 amide~I quanta, each of which has energy $E_0=0.2$ eV.
Because the continuum approximation of Davydov's model \cite{Brizhik1983,Ostrovskaya2001,Chen2017,Vakhnenko2021,Georgiev2020b,Georgiev2022} leads to sech-squared soliton solutions, we have applied the metabolic energy
of a single ATP molecule in the form of $Q=3$
amide~I exciton quanta spread initially over 5 or 7 peptide groups
as a discretized sech-squared pulse given by a discrete set of quantum probability amplitudes
\begin{equation}
\frac{1}{\sqrt{3}}e^{\imath \alpha\omega_{2}}\left\{ A_{3,\alpha} e^{- 3\imath \omega_1},A_{2,\alpha} e^{-2\imath \omega_1},A_{1,\alpha} e^{-\imath \omega_1},A_{0,\alpha},A_{1,\alpha} e^{\imath \omega_1},A_{2,\alpha} e^{2\imath \omega_1},A_{3,\alpha} e^{3\imath \omega_1}\right\} \label{eq:sol}
\end{equation}
with phase factor along the spines $\omega_{1}=\frac{\pi}{12}$ \cite{Georgiev2020b},
phase factor laterally across the spines $\omega_{2}=\frac{2\pi}{3}$ \cite{Brizhik2019,Georgiev2022},
and real amplitudes $A_{i,\alpha}$ that were dependent on the initial spread over peptide groups.
The choice of the initial spread to be over at least 5 peptide groups is justified by the chemical structure and dimensions of the hydrolyzed ATP molecule and the enhanced thermal stability predicted for the resulting solitons \cite{Georgiev2022}.
The discretization procedure of the continuous sech-squared pulse consisted of computing the exact spread for which 95\% of the probability is contained in the target number of peptide groups (5 or 7). Then the infinitely long tails outside the target number of peptide groups were removed, the spatial extent of the remaining trimmed pulse was split into corresponding number of equally wide bins (5 or 7), definite integration was performed to compute the quantum probability for each bin, and the total probability was normalized to~1 by multiplication with~$\frac{100}{95}$.
For the soliton spread over 5 peptide
groups, the non-zero exciton amplitudes were
\begin{equation}
A_{0,\alpha}=\sqrt{0.37},\quad A_{1,\alpha}=\sqrt{0.237},\quad A_{2,\alpha}=\sqrt{0.078}\label{eq:s5}
\end{equation}
whereas for the soliton spread over 7 peptide groups, the non-zero
exciton amplitudes were
\begin{equation}
A_{0,\alpha}=\sqrt{0.232},\quad A_{1,\alpha}=\sqrt{0.199},\quad A_{2,\alpha}=\sqrt{0.126},\quad A_{3,\alpha}=\sqrt{0.059}\label{eq:s7}
\end{equation}
The initial pulse of amide~I energy \eqref{eq:sol} was applied at the N-end of the protein $\alpha$-helix, so that the peptide groups indexed by $n=1$ received non-zero initial exciton amplitudes.
For the soliton spread over 5 groups, the initial exciton amplitudes were populated from \eqref{eq:sol} as follows:
$a_{1,\alpha}(0)= \frac{1}{\sqrt{3}}e^{\imath \alpha\omega_{2}} A_{2,\alpha} e^{- 2\imath \omega_1}$, 
$a_{2,\alpha}(0)= \frac{1}{\sqrt{3}}e^{\imath \alpha\omega_{2}} A_{1,\alpha} e^{- \imath \omega_1}$, 
$a_{3,\alpha}(0)= \frac{1}{\sqrt{3}}e^{\imath \alpha\omega_{2}} A_{0,\alpha}$, 
$a_{4,\alpha}(0)= \frac{1}{\sqrt{3}}e^{\imath \alpha\omega_{2}} A_{1,\alpha} e^{ \imath \omega_1}$,
$a_{5,\alpha}(0)= \frac{1}{\sqrt{3}}e^{\imath \alpha\omega_{2}} A_{2,\alpha} e^{ 2\imath \omega_1}$, 
and $a_{n,\alpha}(0)=0$ for $n\geq 6$.
For the soliton spread over 7 groups, the initial exciton amplitudes were similarly populated from \eqref{eq:sol} starting with:
$a_{1,\alpha}(0)= \frac{1}{\sqrt{3}}e^{\imath \alpha\omega_{2}} A_{3,\alpha} e^{- 3\imath \omega_1}$, 
$a_{2,\alpha}(0)= \frac{1}{\sqrt{3}}e^{\imath \alpha\omega_{2}} A_{2,\alpha} e^{- 2\imath \omega_1}$, etc.
The phonon lattice was initially unperturbed with $b_{n,\alpha}(0)=0$
and $\frac{d}{dt}b_{n,\alpha}(0)=0$.
Reflective boundary conditions were implemented at the two ends of the protein $\alpha$-helix following the procedure by Luo and Piette \cite{Luo2017} by setting $a_{0,\alpha}(t)=a_{41,\alpha}(t)=0$ and $b_{0,\alpha}(t)=b_{41,\alpha}(t)=0$.

Visualization of the soliton trajectory was performed by plotting
the quantum probability of finding the amide~I exciton inside the
$n$th protein unit cell
\begin{equation}
P\left(a_{n}\right)=\sum_{\alpha}|a_{n,\alpha}|^{2}=|a_{n,0}|^{2}+|a_{n,1}|^{2}+|a_{n,2}|^{2}
\end{equation}
Previously, we have shown that bare amide~I exciton pulses were able to self-induce at subpicosecond timescale an accompanying phonon deformation in an initially unperturbed phonon lattice \cite{Georgiev2019a}. This induced localized phonon deformation slows down the motion of the excitons and is referred to as self-trapping effect. In addition, we have found that emitted ``empty'' phonon waves are moving forth and back at the speed of sound in the phonon lattice without ``carrying'' any exciton amplitudes on top \cite{Georgiev2019a}. Because in this present work we are primarily interested in the transport of exciton energy and the 100~ps duration of performed simulations is quite long compared to the time for induction of self-trapping, we consider that as long as the exciton amplitudes remain collected together as a pulse they would be accompanied by a localized phonon deformation.

An assembly of multiple protein subunits could be modeled as a localized
enhancement of the effective mass of the peptide groups in the protein
$\alpha$-helix. A concentric assembly of proteins will lead to ring-like
massive barriers that are distributed equally onto the three $\alpha$-helical
spines. Because large protein subunits contain multiple protein $\alpha$-helices
connected with protein loops, it is expected that for most of the
protein $\alpha$-helices, which are not centrally located, the distribution
of external mass will be excentric. Furthermore, functionally active
protein sites are typically located at the interface of interacting
protein subunits, which also suggests that the excentric distribution
of protein mass onto several of the three $\alpha$-helical spines could
be biologically relevant for the transport and utilization of metabolic
energy by active proteins. To test how concentric or excentric massive
barriers affect the propagation of solitons in protein $\alpha$-helices,
we have increased locally the values for $M_{n,\alpha}$ as multiples
of the average mass $M=1.9\times10^{-25}$~kg of a single amino acid
inside the polypeptide chain.

The total mass of the single protein $\alpha$-helix with $n_{\max}=40$ lattice sites is $3\times40\times1M=120M$. However, functional macromolecular protein complexes, such as actin-tropomyosin-myosin complexes \cite{Behrmann2012}, F$_1$F$_0$-type ATP synthase \cite{Capaldi2002}, voltage-gated ion channels \cite{Catterall1995}, kinesin motors \cite{Liu2012} or SNARE protein complexes \cite{Zhou2015}, have multiple subunits each of which may have several protein $\alpha$-helices.
Furthermore, those protein subunits could be anchored to phosholipid membranes where they are exposed to large mechanical forces or strong electric fields.
For example, 4-$\alpha$-helix bundles of SNARE proteins drive the fusion of synaptic vesicles with diameter of 40 nm \cite{Zhang1998} with the presynaptic plasma membrane, kinesin motors drag huge intracellular cargo vesicles with diameter of 100 nm \cite{Jiang2019} along microtubule tracks, and electrically charged S4 protein $\alpha$-helix sensors inside voltage-gated ion channels experience the physical forces due to strong transmembrane electric field with intensity of $2\times10^7$ V/m \cite{Chanda2008}. Even though the protein interactions are quite complex and not easy to model, for the purposes of the present study we consider that barriers with effective mass of several hundred amino acid masses are biologically feasible. We also would like to note that the computational study is primarily intended to characterize the general physical trends with respect to soliton transmission, reflection or trapping by massive barriers. Consequently, suitable adaptation of the reported results to concrete molecular systems in biological systems would have to take into consideration the relevant structural data with regard of the composition of the molecular system of interest.

The system of equations \eqref{eq:full-1}, \eqref{eq:full-2} has two conserved quantities: the total probability
\begin{equation}
\mathcal{P}(t)=\sum_{n,\alpha}\left|a_{n,\alpha}\right|^{2}=1
\end{equation}
and the expectation value of the gauge transformed total energy
\begin{align}
\mathcal{E}(t) & =\langle\Psi(t)|\hat{H}|\Psi(t)\rangle-QE_{0} -W_{0}\nonumber \\
 & =-\sum_{n,\alpha}\Bigg\{ QJ_{1}\left(a_{n,\alpha}^{*}a_{n+1,\alpha}+a_{n,\alpha}a_{n+1,\alpha}^{*}\right)-QJ_{2}\left(a_{n,\alpha}^{*}a_{n,\alpha+1}+a_{n,\alpha}a_{n,\alpha+1}^{*}\right)\nonumber \\
 & \quad-\frac{1}{2}M_{n,\alpha}\left(\frac{d}{dt}b_{n,\alpha}\right)^{2}-\frac{1}{2}w_{1}\left(b_{n+1,\alpha}-b_{n,\alpha}\right)^{2}-\frac{1}{2}w_{2}\left(b_{n,\alpha+1}-b_{n,\alpha}\right)^{2}\nonumber \\
 & \quad-Q\left[\chi_{l}\left(b_{n,\alpha}-b_{n-1,\alpha}\right)+\chi_{r}\left(b_{n+1,\alpha}-b_{n,\alpha}\right)\right]\left|a_{n,\alpha}\right|^{2}\Bigg\}
\end{align}
where $W_{0}=\langle0_{\textrm{ph}}|\hat{H}_{\textrm{ph}}|0_{\textrm{ph}}\rangle$ is the zero-point energy of the lattice vibrations.
The accuracy with which these two quantities were preserved during the numerical integration of the system of equations of motion was evaluated at the end of the 100~ps simulation period using $\Delta\mathcal{P}=1-\mathcal{P}(100)$ and $\Delta\mathcal{E}=1-\mathcal{E}(100)/\mathcal{E}(0)$. For the simulations with solitons initially spread over 5~peptide groups, the accuracy of preservation was $\Delta\mathcal{P} < 1.67\times10^{-7}$ and $\Delta\mathcal{E} < 1.66\times10^{-7}$, whereas for the simulations with solitons initially spread over 7~peptide groups, it was $\Delta\mathcal{P}< 2.51\times10^{-6}$ and $\Delta\mathcal{E} < 2.75\times10^{-6}$.

\section{Computational results}

\begin{figure}
\begin{centering}
\includegraphics[width=160mm]{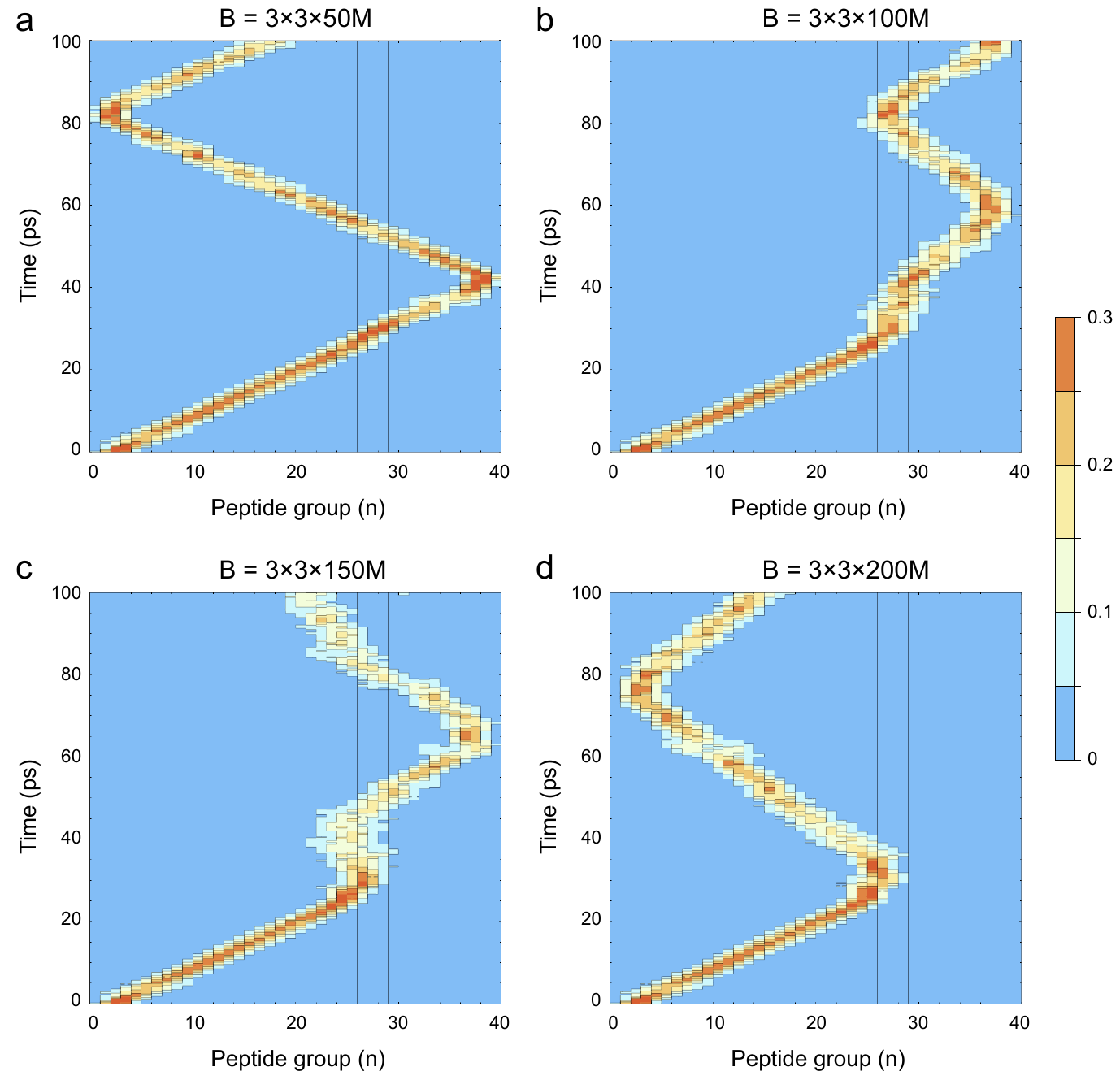}
\par\end{centering}

\caption{\label{fig:1}The quantum dynamics of a molecular soliton initiated by a
sech-squared exciton pulse with $Q=3$ amide~I exciton quanta spread initially over 5 peptide groups
that undergoes tunneling through a concentric massive barrier $B$
centered on all three spines ($3\times$) extending over three peptide
groups ($3\times$), $n$=26--28, each of which with increased effective
mass up to $50M$ (a), $100M$ (b), $150M$ (c) or $200M$ (d). The
soliton is able to tunnel through the lighter barriers whose total
mass is below or equal to $1350M$ (a--c), but is reflected from the
heaviest barrier with total mass of $1800M$ (d). The extent of the
massive barrier along peptide groups $n$=26--28 is indicated with
thin vertical black lines.}
\end{figure}

For establishing a base case for comparison, we have first simulated
the quantum dynamics of a molecular soliton, which was spread initially
over 5 peptide groups, and encountered a concentric massive barrier
on all 3 spines extending over 3 peptide groups, each of which with
the same increased effective mass, $M_{26,\alpha}=M_{27,\alpha}=M_{28,\alpha}\in\{50M,100M,150M,200M\}$
for $\alpha\in\{0,1,2\}$ (Fig.~\ref{fig:1}). The initial speed of
the soliton was 439~m/s before impacting on the massive barrier.
Crucially, when the total mass of the concentric barrier was below or equal to $1350M$,
the soliton tunneled successfully to the other side of the barrier (Fig.~\ref{fig:1}a-c).
The loss of quantum probability following the passage through the barrier increased with the barrier mass (Supplementary Fig.~S1).
For the light concentric barrier with total mass of $450M$,
the maximal probability to detect each exciton on the left side of the barrier was 94.5\% (Supplementary Fig.~S1a)
and the soliton speed remained 439~m/s after the
reflection from the protein $\alpha$-helix end (Fig.~\ref{fig:1}a).
For the heavier concentric barrier with total mass of $900M$, the
soliton remained for 22.6~ps inside the barrier,
the maximal probability to detect each exciton on the left side of the barrier decreased to 90.1\% (Supplementary Fig.~S1b)
and the soliton speed after tunneling through the barrier dropped to 266~m/s (Fig.~\ref{fig:1}b).
Further increasing the total barrier mass to $1350M$ prolonged the time during
which the soliton remained in the vicinity of the concentric barrier,
increased the tunneling delay to 30.3~ps (Fig.~\ref{fig:1}c) and
decreased the maximal probability to detect each exciton on the left side of the barrier to 83.0\% (Supplementary Fig.~S1c).
For the heaviest concentric barrier with total mass of $1800M$, the
soliton was reflected from the barrier with reflection time of 17.9~ps and decreased speed of 274~m/s (Fig.~\ref{fig:1}d). Taken together,
these results confirm that the presence of massive barriers can lead
to compartmentalization of the space available for soliton propagation
within the protein $\alpha$-helix \cite{Georgiev2019b}.

\begin{figure}
\begin{centering}
\includegraphics[width=160mm]{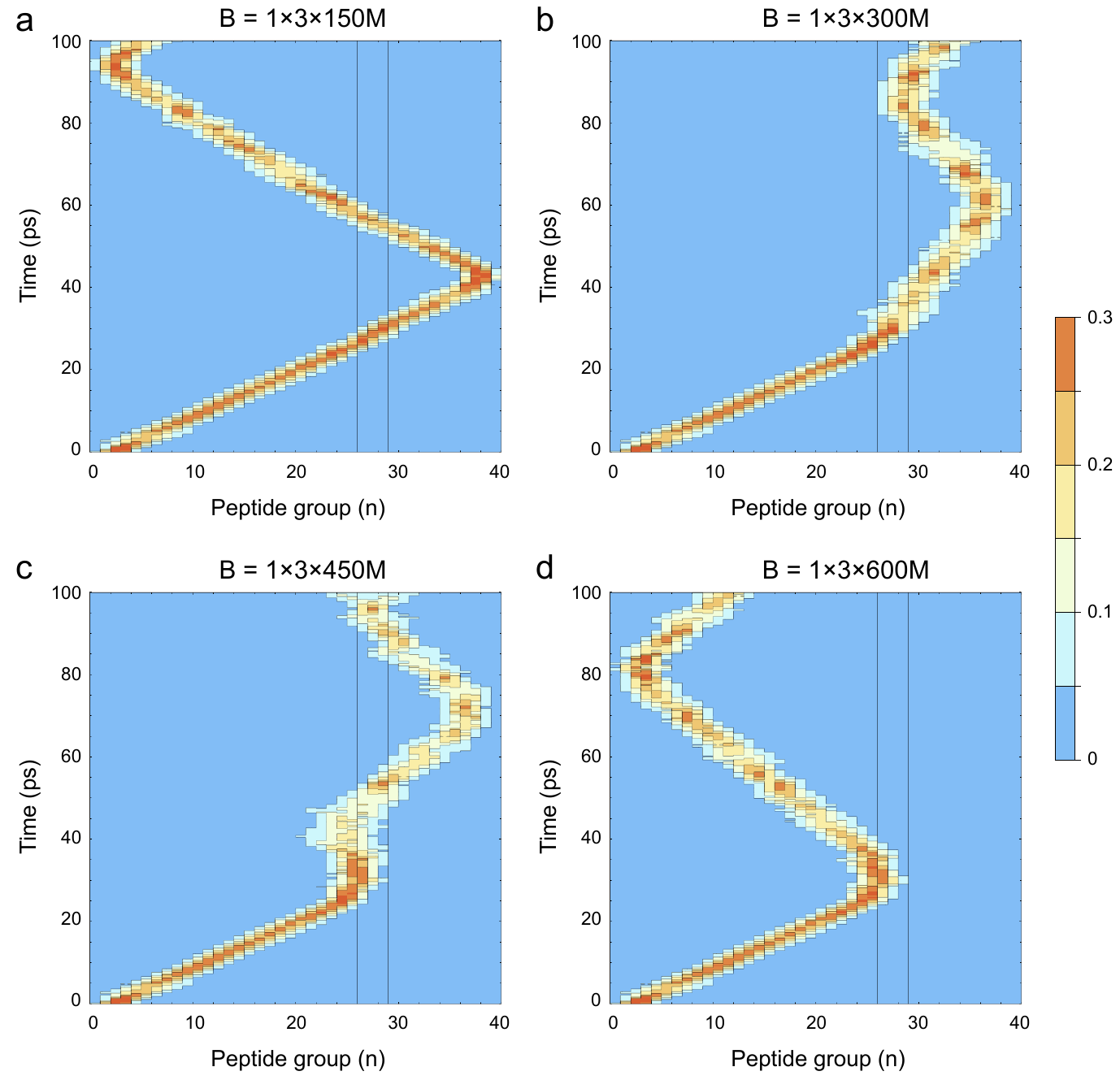}
\par\end{centering}

\caption{\label{fig:2}The quantum dynamics of a molecular soliton initiated by a
sech-squared exciton pulse with $Q=3$ amide~I exciton quanta spread initially over 5 peptide groups
that undergoes tunneling through an excentric massive barrier $B$
placed on a single spine ($1\times$) extending along three peptide
groups ($3\times$), $n$=26--28, each of which with increased effective
mass up to $150M$ (a), $300M$ (b), $450M$ (c) or $600M$ (d). The
soliton is able to tunnel through the lighter barriers whose total
mass is below or equal to $1350M$ (a--c), but is reflected from the
heaviest barrier with total mass of $1800M$ (d). The extent of the
massive barrier along peptide groups $n$=26--28 is indicated with
thin vertical black lines.}
\end{figure}

Next, we consider the cumulative effect of all masses present
in the previous concentric barrier onto a single spine, thereby producing
a massive excentric barrier, $M_{26,\alpha}=M_{27,\alpha}=M_{28,\alpha}\in\{150M,300M,450M,600M\}$
for $\alpha=0$ and $M_{26,\alpha}=M_{27,\alpha}=M_{28,\alpha}=1M$
for $\alpha\in\{1,2\}$ (Fig.~\ref{fig:2}). The initial speed of
the soliton remained 439~m/s before impacting on the massive excentric
barrier. When the total mass of the excentric barrier was below or
equal to $1350M$, the soliton was able to tunnel again successfully
to the other side of the barrier (Fig.~\ref{fig:2}a-c).
For the light excentric barrier with total mass of $450M$,
the maximal probability to detect each exciton on the left side of the barrier was 95.5\% (Supplementary Fig.~S2a)
and the soliton speed dropped to 309~m/s after the second passing through the excentric barrier
(Fig.~\ref{fig:2}a). For the heavier excentric barrier with total
mass of $900M$, the soliton remained for 18.5~ps inside the barrier,
the maximal probability to detect each exciton on the left side of the barrier decreased to 90.3\% (Supplementary Fig.~S2b),
and the speed after tunneling through the barrier dropped to 160~m/s (Fig.~\ref{fig:2}b).
Further increasing the total barrier mass to $1350M$
prolonged the time during which the soliton remained in the vicinity
of the excentric barrier, increased the tunneling delay to 34.3~ps (Fig.~\ref{fig:2}c) and
decreased the maximal probability to detect each exciton on the left side of the barrier to 80.0\% (Supplementary Fig.~S2c).
For the heaviest excentric barrier with total
mass of $1800M$, the soliton was reflected from the barrier with
reflection time of 18.2~ps and decreased speed of 244~m/s (Fig.~\ref{fig:2}d).
Taken together, these results indicate that the collection of effective
mass onto a single spine in the barrier site is able to enhance the
interaction time between the soliton and the barrier, which in turn
leads to spreading of the soliton and decreasing its speed.

\begin{figure}
\begin{centering}
\includegraphics[width=160mm]{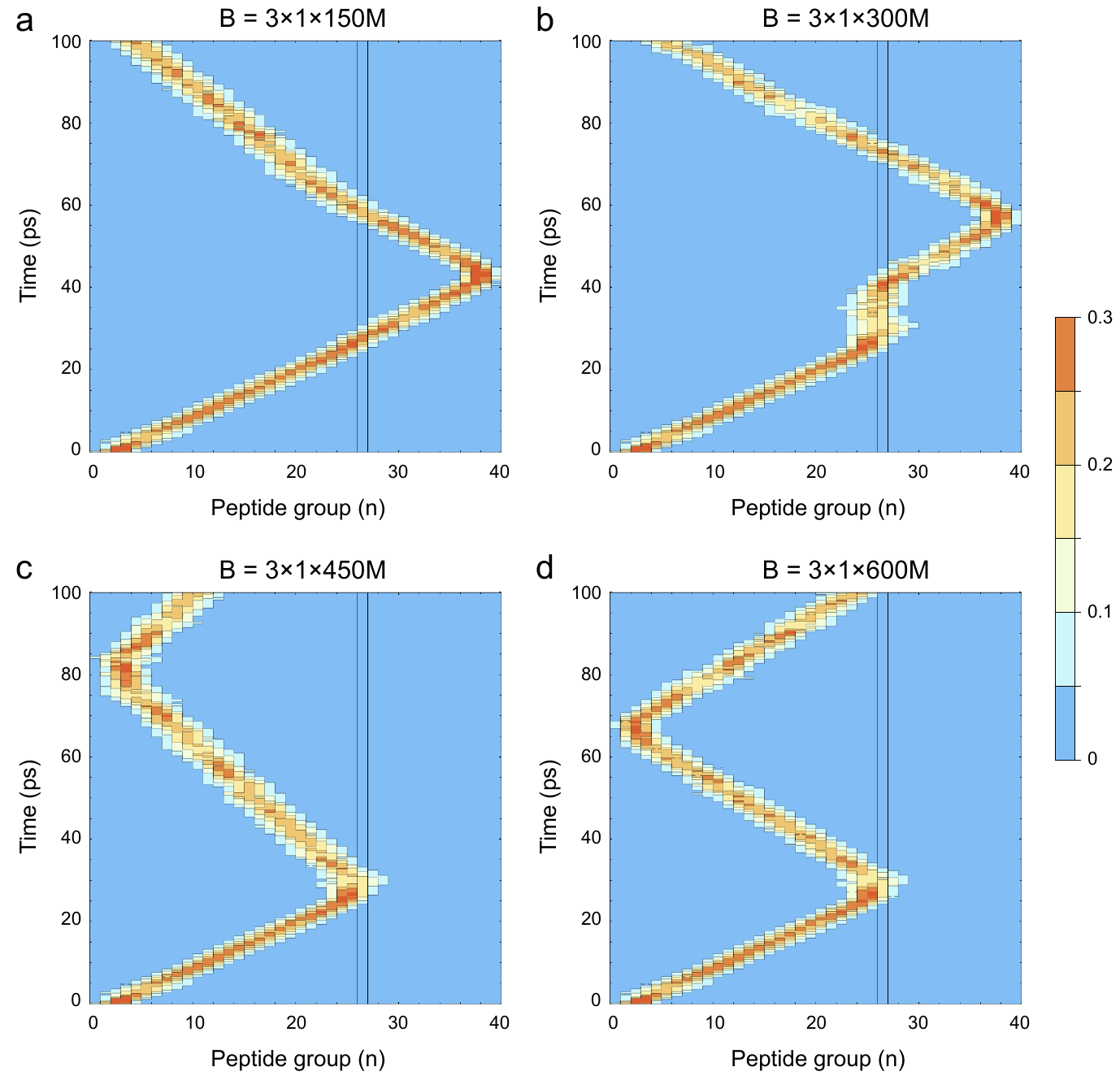}
\par\end{centering}

\caption{\label{fig:3}The quantum dynamics of a molecular soliton initiated by a
sech-squared exciton pulse with $Q=3$ amide~I exciton quanta spread initially over 5 peptide groups
that undergoes tunneling through a concentric massive barrier $B$
centered on all three spines ($3\times$) extending along a single
peptide group ($1\times$), $n=26$, with increased effective mass
up to $150M$ (a), $300M$ (b), $450M$ (c) or $600M$ (d). The soliton
is able to tunnel through the lighter barriers whose total mass is
below or equal to $900M$ (a-b), but is reflected from the heavier
barriers with total mass greater or equal to $1350M$ (c-d). The extent
of the massive barrier along the peptide group $n=26$ is indicated
with thin vertical black lines.}
\end{figure}

The effect of concentric collection of the barrier mass onto a ring
extending over a single peptide group, $M_{26,\alpha}\in\{150M,300M,450M,600M\}$
for $\alpha\in\{0,1,2\}$ (Fig.~\ref{fig:3}) differed in significant
ways from the soliton dynamics observed with the wider barriers (Figs.~\ref{fig:1} and \ref{fig:2}).
Firstly, the soliton was able to tunnel successfully to the other side of the barrier only when the total
mass of the narrow concentric barrier was below or equal to $900M$ (Fig.~\ref{fig:3}a-b and Supplementary Fig.~S3a-b), but reflected from the barrier when the total mass was greater or equal to $1350M$ (Fig.~\ref{fig:3}c-d and Supplementary Fig.~S3c-d). Secondly, the
soliton was able to tunnel through the barrier for a second time in the
reverse direction for the narrow concentric barrier with total mass
of $900M$ (Fig.~\ref{fig:3}b), whereas it remained in the left compartment
of the protein $\alpha$-helix for the wider concentric barrier with
total mass of $900M$ (Fig.~\ref{fig:1}b) or the wider excentric
barrier with total mass of $900M$ (Fig.~\ref{fig:2}b). Thirdly, for
the heaviest narrow concentric barrier with total mass of $1800M$,
the soliton was reflected from the barrier with shorter reflection
time of 11.6~ps (Fig.~\ref{fig:3}d) compared to the respective reflection
times of 17.9~ps and 18.2~ps from the wider barriers (Figs.~\ref{fig:1}d and \ref{fig:2}d). Fourthly, the speed of the reflected soliton was
much higher, 329~m/s, after bouncing from the heaviest narrow concentric
barrier with total mass of $1800M$ (Fig.~\ref{fig:3}d) compared
to the respective soliton speeds of 274~m/s and 244~m/s after reflection
from the wider barriers (Figs.~\ref{fig:1}d and \ref{fig:2}d).
Combining all of these results indicates that the narrower concentric barrier
increases the probability for a soliton reflection, and reduces the interaction
time between the soliton and the barrier, which in turn leads to a lesser loss of speed for the reflected soliton.

Solitons that are spread over more peptide groups have lower energy
and move at lower speeds compared to solitons that are spread over
fewer peptide groups \cite{Georgiev2019a,Georgiev2020a}. From the
multinomial theorem \cite{Tauber1963} and the action of bosonic creation
or annihilation operators on the vacuum state \cite{Fock1932,Faddeev2004,Gross2004},
the expectation value of the exciton energy operator is found to be
\begin{align}
\langle\psi_{\textrm{ex}}(t)|\hat{H}_{\textrm{ex}}|\psi_{\textrm{ex}}(t)\rangle & =Q\sum_{n,\alpha}\left[E_{0}|a_{n,\alpha}|^{2}-J_{1}\left(a_{n,\alpha}^{*}a_{n+1,\alpha}+a_{n,\alpha}a_{n+1,\alpha}^{*}\right)+J_{2}\left(a_{n,\alpha}^{*}a_{n,\alpha+1}+a_{n,\alpha}a_{n,\alpha+1}^{*}\right)\right]\\
 & =QE_{0}-2QJ_{1}\sum_{n,\alpha}\left[\textrm{Re}\left(a_{n,\alpha}\right)\textrm{Re}\left(a_{n+1,\alpha}\right)+\textrm{Im}\left(a_{n,\alpha}\right)\textrm{Im}\left(a_{n+1,\alpha}\right)\right]\nonumber \\
 & \quad+2QJ_{2}\sum_{n,\alpha}\left[\textrm{Re}\left(a_{n,\alpha}\right)\textrm{Re}\left(a_{n,\alpha+1}\right)+\textrm{Im}\left(a_{n,\alpha}\right)\textrm{Im}\left(a_{n,\alpha+1}\right)\right]
\end{align}
For a detailed calculation of the expectation values of two-site exciton operators, see \ref{app-A}.

\begin{figure}
\begin{centering}
\includegraphics[width=160mm]{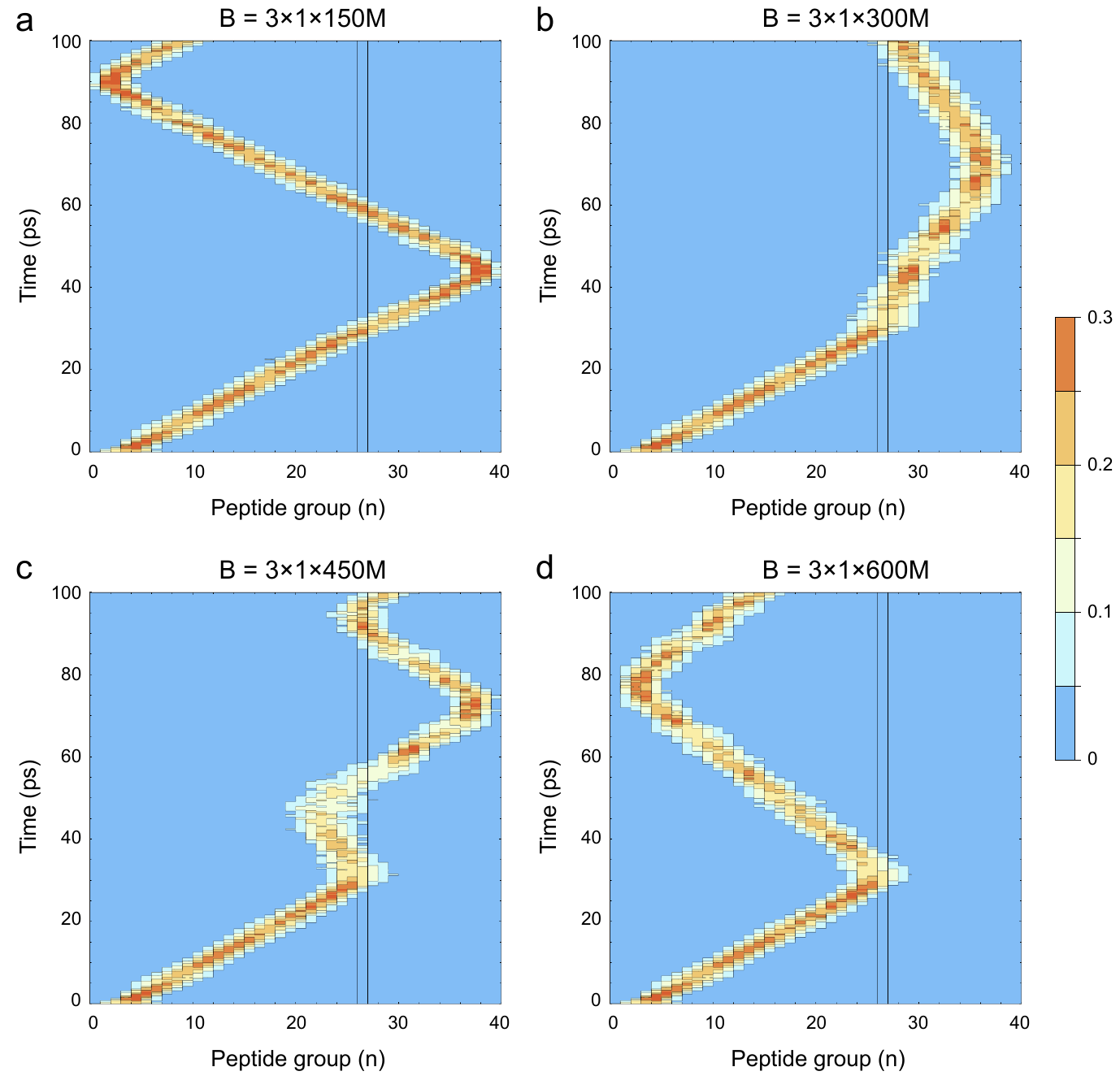}
\par\end{centering}

\caption{\label{fig:4}The quantum dynamics of a molecular soliton initiated by a
sech-squared exciton pulse with $Q=3$ amide~I exciton quanta spread initially over 7 peptide groups
that undergoes tunneling through a concentric massive barrier $B$
centered on all three spines ($3\times$) extending along a single
peptide group ($1\times$), $n=26$, with increased effective mass
up to $150M$ (a), $300M$ (b), $450M$ (c) or $600M$ (d). The soliton
is able to tunnel through the lighter barriers whose total mass is
below or equal to $1350M$ (a--c), but is reflected from the heaviest
barrier with total mass of $1800M$ (d). The extent of the massive
barrier along the peptide group $n=26$ is indicated with thin vertical
black lines.}
\end{figure}

For the soliton \eqref{eq:sol} that is spread over 5~peptide
groups as in \eqref{eq:s5}, the initial exciton energy is
\begin{equation}
\langle\hat{H}_{\textrm{ex}}\rangle\approx 3 \left(E_{0}-1.67J_{1}-J_{2}\right)
\end{equation}
When the soliton \eqref{eq:sol} is spread over 7~peptide groups as
in \eqref{eq:s7}, the initial exciton energy is decreased to
\begin{equation}
\langle\hat{H}_{\textrm{ex}}\rangle\approx 3 \left(E_{0}-1.78J_{1}-J_{2}\right)
\end{equation}
and the initial speed of the soliton drops to 365~m/s (Fig.~\ref{fig:4}).
The latter wider soliton also tunnels readily through the lighter
narrow concentric barriers with total mass of $450M$ (Fig.~\ref{fig:4}a)
or total mass of $900M$ (Fig.~\ref{fig:4}b). In contrast to the
narrow soliton spread over 5 peptide groups, which reflects from the
narrow concentric barrier with total mass of $1350M$ (Fig.~\ref{fig:3}c),
however, the wider soliton spread over 7 peptide groups successfully
tunnels through the barrier (Fig.~\ref{fig:4}c and Supplementary Fig.~S4c). Consistent with
stronger interaction with the massive barrier, the wider soliton spread
over 7~peptide groups is reflected from the heaviest narrow concentric
barrier with total mass of $1800M$ with increased reflection time
to 12.4~ps (Fig.~\ref{fig:4}d) compared to 11.6~ps observed for
the narrower soliton spread over 5~peptide groups (Fig.~\ref{fig:3}d).
Taken together, these results show that the efficacy of massive barriers
to transmit or reflect incoming solitons depends critically on the
soliton energy. Due to their lower energy, wider solitons manifest
increased tunneling probability (Supplementary Fig.~S4c), stronger interaction with the massive
barriers and increased reflection times. This provides biological
systems with at least two distinct strategies for exerting control
over the transport of energy inside proteins: one strategy is to control
the effective mass of the barriers due to interacting proteins through
protein subunit assembly, and a second strategy is to control the
initial soliton spread through biochemical regulation of the release
of metabolic energy from ATP at protein hydrolytic sites.

\begin{figure}
\begin{centering}
\includegraphics[width=160mm]{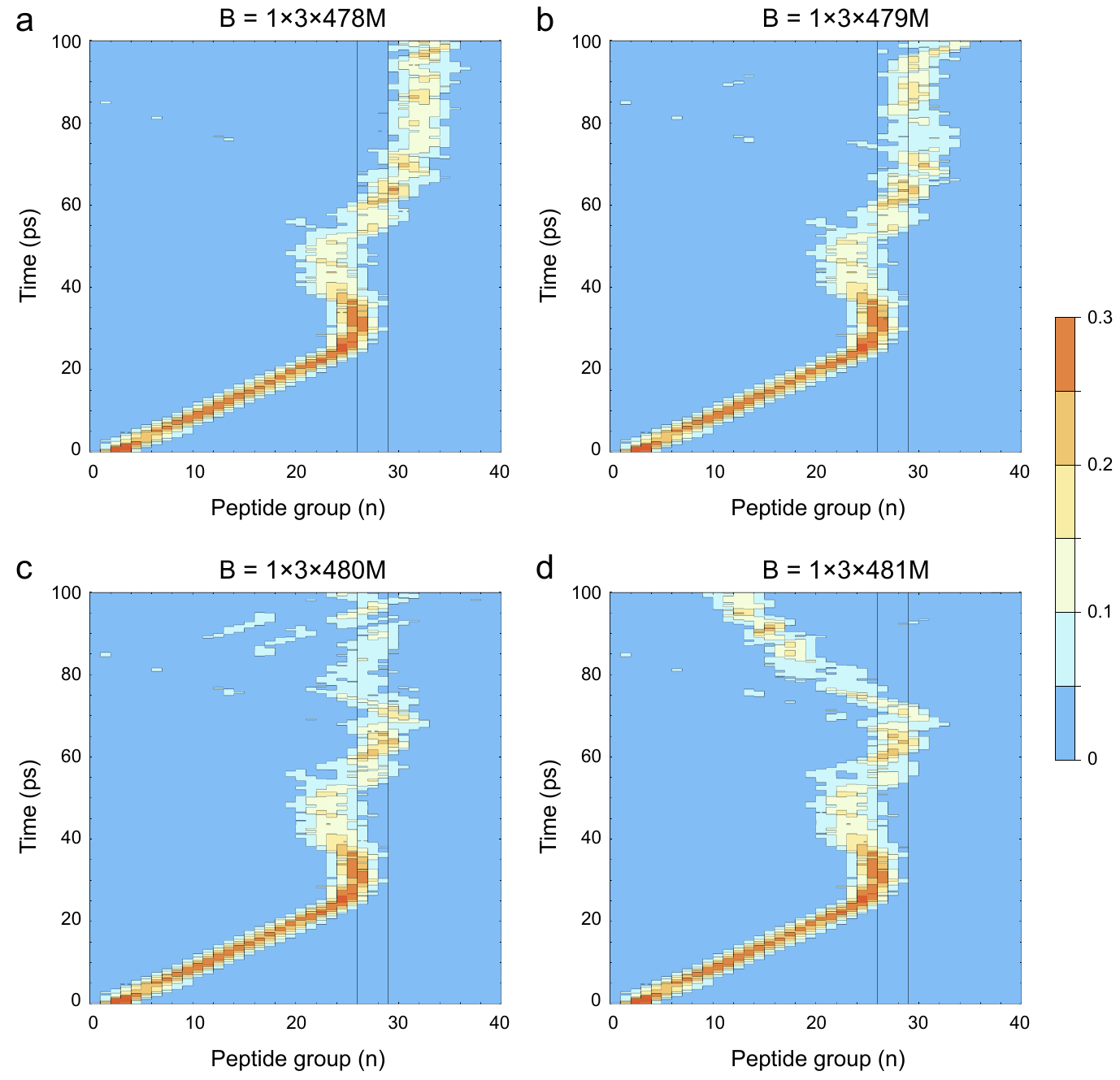}
\par\end{centering}

\caption{\label{fig:5}The trapping of a molecular soliton initiated by a sech-squared
exciton pulse with $Q=3$ amide~I exciton quanta spread initially over 5 peptide groups that impacts
upon an excentric massive barrier $B$ placed on a single spine ($1\times$)
extending along three peptide groups ($3\times$), $n$=26--28, each
of which with increased effective mass up to $478M$ (a), $479M$
(b), $480M$ (c) or $481M$ (d). The soliton is able to tunnel through
the lighter barriers whose total mass is below or equal to $1437M$
(a--b), but remains in the vicinity on the right side of the barrier
until the end of the simulation period of 100~ps. The soliton remains
trapped inside the barrier whose total mass is 1440M (c) and is reflected
from the heaviest barrier with a total mass of $1443M$ (d). The extent
of the massive barrier along peptide groups $n$=26--28 is indicated
with thin vertical black lines.}
\end{figure}

\begin{figure}
\begin{centering}
\includegraphics[width=160mm]{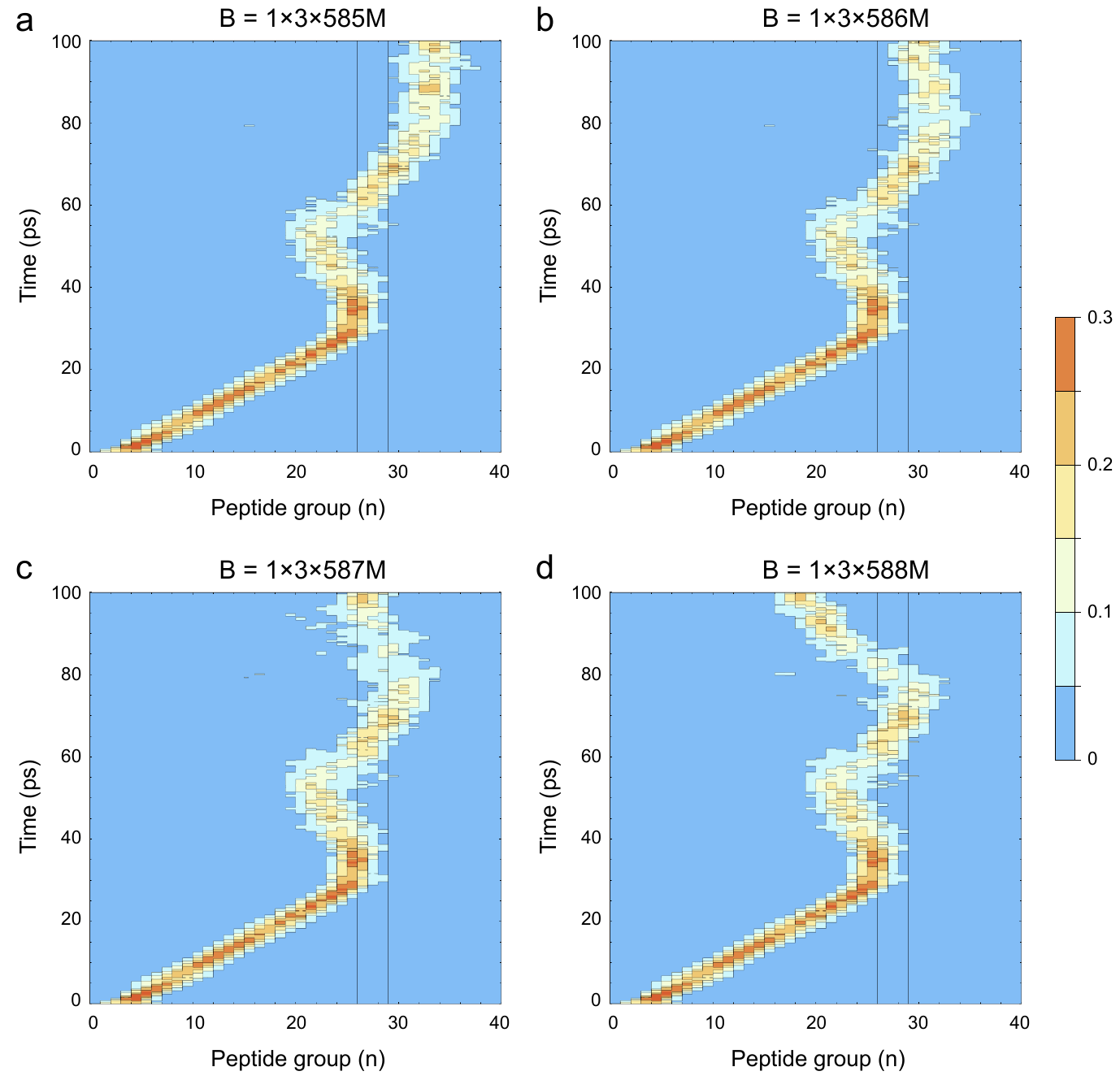}
\par\end{centering}

\caption{\label{fig:6}The trapping of a molecular soliton initiated by a sech-squared
exciton pulse with $Q=3$ amide~I exciton quanta spread initially over 7 peptide groups that impacts
upon an excentric massive barrier $B$ placed on a single spine ($1\times$)
extending along three peptide groups ($3\times$), $n$=26--28, each
of which with increased effective mass up to $585M$ (a), $586M$
(b), $587M$ (c) or $588M$ (d). The soliton is able to tunnel through
the lighter barriers whose total mass is below or equal to $1758M$
(a--b), but remains in the vicinity on the right side of the barrier
until the end of the simulation period of 100~ps. The soliton remains
trapped inside the barrier whose total mass is $1761M$ (c) and is
reflected from the heaviest barrier with a total mass of $1764M$ (d).
The extent of the massive barrier along peptide groups $n$=26--28
is indicated with thin vertical black lines.}
\end{figure}

Lastly, we have explored the capacity of excentric massive barriers
to trap the solitons at the interface of interacting protein subunits,
which would provide a physical mechanism for utilization of the soliton
energy to do useful work at functionally active protein sites. Having
observed that the soliton tunneling is replaced with a soliton reflection
as the barrier mass grows (Figs.~\ref{fig:1}--\ref{fig:4}), we conjecture the existence of some barrier mass for which the excentric
barrier located at one of the protein $\alpha$-helix spines is able
to trap the soliton inside the barrier region. Indeed, we have confirmed
that the narrow soliton spread over 5 peptide groups is trapped by
an excentric massive barrier with total mass of $1440M$ (Fig.~\ref{fig:5} and Supplementary Fig.~S5),
whereas the wider soliton spread over 7 peptide groups is trapped
by an excentric barrier whose total mass is $1761M$ (Fig.~\ref{fig:6} and Supplementary Fig.~S6).
These results corroborate the observed higher tunneling probability
for wider solitons (Figs.~\ref{fig:3} and \ref{fig:4}) and demonstrate
that higher overall effective barrier mass is required to nullify
the tunneling probability for the wider soliton. The range of barrier
masses, for which the soliton remains trapped in the vicinity of the
barrier for a prolonged period of time, corresponds to 5--10 amino
acid masses. The latter fact is particularly useful from a biological
perspective because the amino acid composition of proteins could be
controlled in living cells through mRNA splicing \cite{Wang2008,Chen2009,Nilsen2010,Shi2017}.

The mechanical forces that are exerted upon actin-tropomyosin-myosin complexes during skeletal muscle contraction \cite{Caremani2015} or upon zipping SNARE protein complexes during fusion of synaptic vesciles \cite{Min2013} could increase significantly the effective mass of portions of protein $\alpha$-helices located at protein-protein attachment interfaces or protein-lipid membrane insertion points. Although the effective masses of barriers imposed onto individual protein $\alpha$-helices need to be evaluated for each functional macromolecular protein complex on a case by case basis, here we consider that the molecular mass of the myosin heavy chain, which contains 1959 amino acids in its sequence \cite{Shohet1989}, is able to provide a meaningful upper bound of about $2000M$ on the total barrier mass that can act upon a target protein $\alpha$-helix.
The presence of isolated point impurities has been previously shown to reduce the amplitude of impacting solitons through emission of ``quasi-particles'' in the continuum approximation \cite{Kivshar1987,Kivshar1989}. Here, we have shown that for the biologically relevant parameter values of the discrete Davydov system, the solitons can impact upon very massive barriers with total mass of up to $1800M$ and suffer only a minimal loss of exciton quantum probability by the end of the 100~ps simulation period (see Supplementary Figs.~S1-S6). Because in the majority of macromolecular protein systems it is expected that the effective barrier masses are lower than  $1800M$, our computational results suggest that biological massive barriers modulate the dynamics of protein solitons without destroying them.

\begin{figure}
\begin{centering}
\includegraphics[width=160mm]{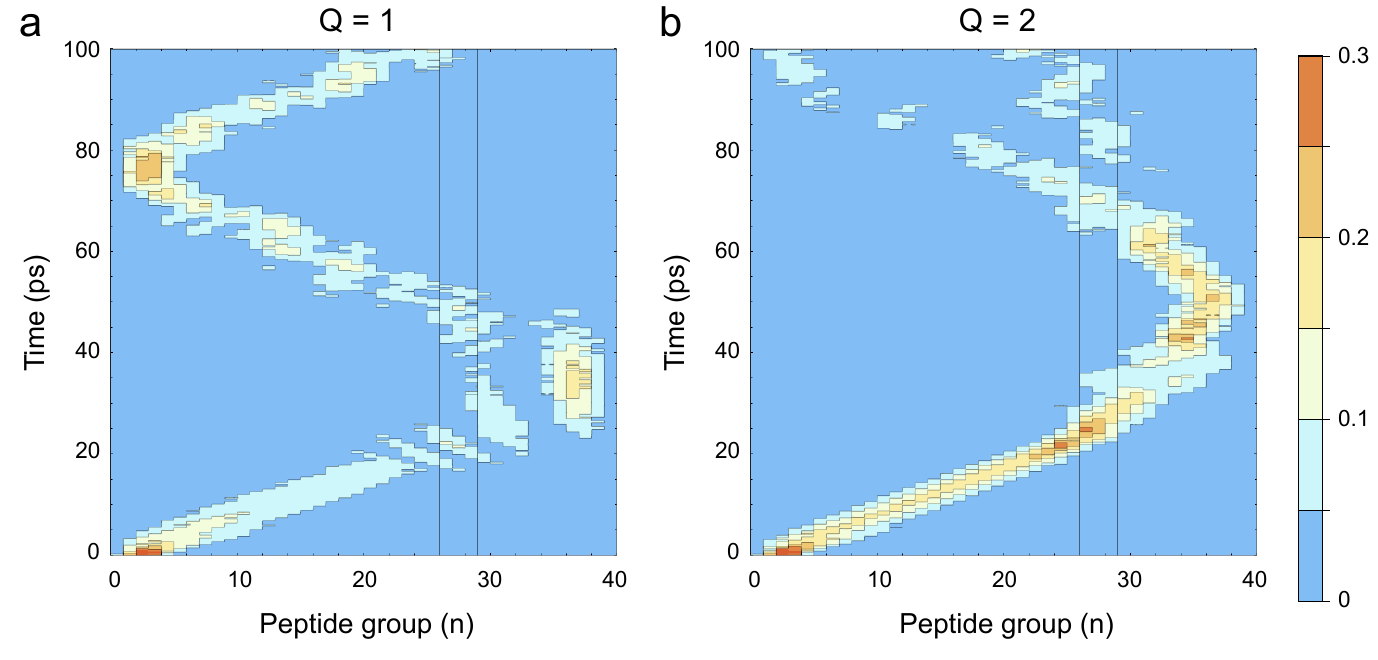}
\par\end{centering}

\caption{\label{fig:7}The quantum dynamics of a molecular soliton initiated by a
sech-squared exciton pulse with small number of amide~I exciton quanta, $Q=1$ (a) or $Q=2$ (b), spread initially over 5 peptide groups
that undergoes tunneling through a concentric massive barrier $B$
centered on all three spines ($3\times$) extending over three peptide
groups ($3\times$), $n$=26--28, each of which with increased effective mass up to $100M$.
The massive barrier is able to disperse the soliton when the number of quanta is $Q\leq 2$ (a,b), however, there is only a minor loss of exciton quantum probability when $Q=3$ (see Fig.~\ref{fig:1}b).}
\end{figure}

It is worth emphasizing that the apparent enhanced stability of the molecular solitons that we report is due to the cooperative action of $Q=3$ amide~I exciton quanta \cite{Georgiev2022}. In older literature, as reviewed by Scott \cite{Scott1992}, it is incorrectly assumed that the energy released from the hydrolysis of a single ATP molecule can excite at most $Q=2$ amide~I exciton quanta. Recent experimental measurements, however, establish that the hydrolysis of a single ATP molecule suffices to excite $Q=3$ amide~I exciton quanta \cite{Barclay2020,Kammermeier1982,Weiss2005}.
Here, we have simulated quantum dynamics of molecular soliton initiated by a
sech-squared exciton pulse spread initially over 5 peptide groups and have confirmed that the impact onto a massive barrier with total mass of $900M$ could indeed disperse the soliton for $Q\leq 2$ (Figs.~\ref{fig:7}a,b), but preserves the soliton largely intact for $Q= 3$ (Fig.~\ref{fig:1}b).
The stabilization of the soliton by increased number of amide~I exciton quanta $Q$ is explained by the stronger nonlinear exciton--phonon coupling, due to the fact that $Q$ acts as a multiplicative factor of $\chi_r$ and $\chi_l$ in the quantum equations of motion \eqref{eq:full-1} and \eqref{eq:full-2}.
This also contextualizes previous reports by Kivshar and Malomed \cite{Kivshar1987,Kivshar1989} on the stability of Davydov solitons in the presence of impurities, namely, the dispersal of the solitons by massive barriers is pronounced only when the nonlinearity of the system is weak, as estimated by the product of $Q$ and $\chi_r$ or $\chi_l$.

The purely quantum derivation of the equations of motion \eqref{eq:full-1} and \eqref{eq:full-2} emphasizes the non-classical nature of the simulated molecular solitons. From the generalized Ehrenfest theorem \cite{Georgiev2019a} and the assumed validity of the adiabatic approximation \cite{Kerr1987,Kerr1990,Zolotaryuk1988,Georgiev2020a,Georgiev2022}, it follows that the quantum state vector $|\Psi(t)\rangle$ of the composite system is uniquely reconstructed from the numerically simulated $a_{n,\alpha}$, $b_{n,\alpha}$ and $\frac{d}{dt} b_{n,\alpha}$ as follows
\begin{equation}
|\Psi(t)\rangle=\frac{1}{\sqrt{Q!}}\left[\sum_{n,\alpha}a_{n,\alpha}(t)\hat{a}_{n,\alpha}^{\dagger}\right]^{Q}|0_{\textrm{ex}}\rangle e^{-\frac{\imath}{\hbar}\sum_{n',\alpha'}\left(b_{n',\alpha'}(t)\hat{p}_{n',\alpha'}-M_{n',\alpha'}\left[\frac{d}{dt}b_{n',\alpha'}(t)\right]\hat{u}_{n',\alpha'}\right)}|0_{\textrm{ph}}\rangle
\end{equation}
This state satisfies a large number of quantum information-theoretic theorems and the quantum uncertainty principle \cite{Georgiev2022b}.
Consequently, the plotted exciton quantum probability amplitudes should be interpreted only as expectation values of possible exciton position measurements, and not as trajectories of classical objects with fixed positions and velocities.
In order to highlight the quantum tunneling of exciton quantum probability amplitudes through the massive barriers in the protein $\alpha$-helix, we also deliver Supplementary Videos 1--3 for each simulation shown in Figs.~\ref{fig:7}a,b or \ref{fig:1}b.
Because the moduli of the exciton quantum probability amplitudes are the square roots of the corresponding quantum probabilities, the dispersive effects of the massive barriers upon the molecular soliton for $Q\leq 2$ (Supplementary Videos~1--2) are visually magnified in the videos as compared to the contour plots.
The soliton stabilization effect due to increased number of amide~I excition quanta $Q=3$ is also visibly appreciable (Supplementary Video~3), however, for proper estimation of the efficiency of energy transport through molecular solitons one needs to compute the loss of quantum probability rather than the loss of quantum probability amplitude.

In order to obtain an exact quantitative measure for the loss of exciton quantum probability by the soliton for different values of $Q$, we have divided the length of the protein $\alpha$-helix into 36 overlapping regions with length of 5 peptide groups. Then, at $t=100$~ps we have computed the loss as
\begin{equation}
1-\max\left\{ \sum_{n=i+1}^{i+5}\sum_{\alpha=0}^{2}\left|a_{n,\alpha}(t)\right|^{2}\right\} _{i=0}^{35}
\end{equation}
The lost exciton probability was 59.7\% for $Q=1$ (Fig.~\ref{fig:7}a), 63.6\% for $Q=2$ (Fig.~\ref{fig:7}b), and only 15.3\% for $Q=3$ (Fig.~\ref{fig:1}b).
An alternative, less stringent approach to evaluate the lost exciton probability could allow for certain amount of widening of the soliton that is initially collected onto the length of 5 peptide groups. For example, division of the protein $\alpha$-helix length into 34 overlapping regions with length of 7 peptide groups and computing
\begin{equation}
1-\max\left\{ \sum_{n=i+1}^{i+7}\sum_{\alpha=0}^{2}\left|a_{n,\alpha}(t)\right|^{2}\right\} _{i=0}^{33}
\end{equation}
reduces the estimated loss of exciton probability to 52.3\% for $Q=1$, 59.9\% for $Q=2$ and 14.6\% for $Q=3$.
According to Davydov ``for describing real systems even unstable solitary waves can be significant if their lifetime is long in comparison with the time during which the phenomenon under study takes place'' \cite{Davydov1986b}.
In line with the latter statement, throughout this work we have used the term ``soliton'' in a general descriptive sense to refer to localized waves of amide~I excitons that avoid dispersal of half their initial probability for a physiologically meaningful period of time of at least 30-40~ps during which the whole extent of the protein $\alpha$-helix could be traversed.

The reflectivity of the massive barriers increases with the increase of the total barrier mass and the decrease of the barrier width. From the simulated performance of extreme barrier cases, with upper bound of $1800M$ for the total barrier mass and lower bound of $n=1$ for the barrier width, we have found that the solitons lose only a minimal amount of exciton quantum probability. Because biological barriers are expected to be wider and less massive, our numerical results imply that the protein solitons are quite robust to local increases of the effective mass and preserve their amplitudes for physiologically relevant periods of times extending over hundreds of picoseconds. On the other hand, we have also demonstrated that any local increase of the effective mass slows down the motion of the soliton and increases the time spent by the excitons in the region inside or near the massive barrier.

Given the complicated three-dimensional structure of macromolecular protein
complexes, it is not expected that biological systems operate by imposing
perfectly concentric barriers. Our computer simulations, however,
have demonstrated that excentric massive barriers could also achieve
the same level of performance with respect to soliton tunneling, trapping
or reflection, provided that the total mass of the excentric barrier
is similar to the total mass of the concentric barrier.
This grants living systems a versatile physical mechanism in order to compartmentalize
metabolic energy, and to utilize that at functionally active protein sites
for the purpose of implementing the essential work.

\section{Discussion}

Conceptually, the physical processes that support protein function could be allocated into three stages: (i) ATP hydrolysis inside a protein ATP hydrolytic site releases free energy, (ii) transport of the free energy from ATP hydrolytic site to protein active site, and (iii) utilization of the free energy at the protein active site for doing work, namely, execution of some particular biological function. Typically, the ATP hydrolytic site and the protein active site do not coincide for macromolecular protein complexes, which implies that biological systems should have evolved mechanisms that transport the energy within the protein in a manner that maximizes the probability that the energy is utilized at the protein active site. Here, we explore the possibility that the free energy could be transported along a protein $\alpha$-helix, which has one of its ends located near the ATP hydrolytic site and has the other end near the protein active site. For the purposes of the present study, we will take for granted that the ATP hydrolysis is able to excite the amide~I oscillators (C=O~bond stretching) in the nearby receptive protein $\alpha$-helix and subsequently the protein active site is able to utilize the amide~I energy for doing work. Then, the main research question that remains to be answered is whether there exists a physical mechanism that can prevent the soliton from bouncing forth-and-back between the two ends of the protein $\alpha$-helix, but instead prolongs the relative amount of time during which the soliton remains in the vicinity of the protein active site where it can be absorbed and its energy converted into useful work.

The assembly of protein subunits into functional complexes introduces
intersubunit interactions \cite{Walker2009,Staudt2011,Hilton2012},
which may lead to the local enhancement of the effective mass of peptide
groups \cite{Georgiev2019b,Georgiev2020a}. The existence of such
massive barriers in protein $\alpha$-helices is expected to affect
the transportation, delivery and utilization of energy carried by
three-spine solitons. In this work, we have investigated the differential
effects upon quantum dynamics of three-spine solitons in the presence
of concentric or excentric massive barriers located on all, or on some,
of the protein $\alpha$-helix spines.

We observed that the presence of massive barriers is conducive to compartmentalizing
the space available for soliton propagation within the protein $\alpha$-helix.
The total mass of the effective massive barriers could be regulated by attachment of different regulatory or auxiliary subunits to the main protein complex \cite{Chorev2015} or through additional attachment of specialized protein clamps \cite{Shon2018,Georgiev2018}.
For lighter barriers, the soliton may tunnel to the other side of
the barrier with picosecond time delay, whereas for heavy barriers
the soliton becomes reflected from the barrier.
The action of excentric barriers upon soliton dynamics is comparable to that by concentric
barriers with similar total effective mass.
For a certain range of barrier masses, the soliton may remain permanently trapped either
inside, or in the vicinity of the massive barrier. Because protein
active sites are typically located at the interface of interacting
subunits where effective mass is locally increased, the soliton trapping
provides a physical mechanism for utilization of soliton energy for
the performance of biologically useful work at protein active sites.

The ATP hydrolytic sites, which are three-dimensional pockets inside the protein, are different in different protein complexes \cite{Guo2022,Weber2000,Devi2015,Vorobiev2003,Cross2016}. Different ATP hydrolytic sites will have different three-dimensional geometry and will be located at different distance and angle from the receptive protein $\alpha$-helix that receives the ATP released energy for subsequent soliton transport. Naturally, different distances and angles will produce different spreads of the initial exciton energy pulse and these will result in solitons with different widths in different protein complexes. Here, we take it for granted that evolution through natural selection will optimize the geometry of the ATP hydrolytic sites in different protein complexes so that the spreads of the initial exciton energy pulses are best suited for the corresponding protein functions \cite{Georgiev2020a}.
Comparative analysis of the presented numerical simulations has demonstrated that
solitons of greater width move at lower speeds, tunnel
through heavier barriers, and require greater total barrier mass in order to be trapped. 
This highlights the importance of organization
of ATP hydrolytic sites in active proteins for targeted release of
metabolic energy into pulses with appropriate width. The two distinct
physical mechanisms for soliton trapping, either through control of
the soliton width or the total barrier mass, elucidate the physiological
possibilities for regulating protein function in living systems.

\section*{CRediT authorship contribution statement}

\textbf{Danko D. Georgiev}: Writing -- review \& editing, Writing -- original draft, Visualization, Validation, Software, Resources, Project administration, Methodology, Investigation, Formal analysis, Data curation, Conceptualization.
\textbf{James F. Glazebrook}: Writing -- review \& editing, Supervision, Methodology, Investigation, Formal analysis, Conceptualization.

\section*{Declaration of competing interest}

The authors declare that they have no known competing financial
interests or personal relationships that could have appeared to influence
the work reported in this paper.

\section*{Data availability}

No data was used for the research described in the article.

\section*{Acknowledgments}

We would like to thank four anonymous reviewers for their valuable feedback, which helped us improve the presentation of our work.

\appendix

\section{Expectation values of two-site exciton operators}
\label{app-A}

The normalization of quantum probability $\mathcal{P}(t)=\sum_{n,\alpha}\left|a_{n,\alpha}\right|^{2}=1$
in the restricted Hartree--Fock ansatz~\eqref{eq:Hartree} allows for direct
calculation of the expectation value of two-site exciton operators
such as $\hat{a}_{1}^{\dagger}\hat{a}_{2}$ with the use of the multinomial
theorem as follows
\begin{align}
\langle\hat{a}_{1}^{\dagger}\hat{a}_{2}\rangle & =\frac{1}{Q!}\left\langle 0_{\textrm{ex}}\right|\left[\sum_{q}a_{q}^{*}\hat{a}_{q}\right]^{Q}\hat{a}_{1}^{\dagger}\hat{a}_{2}\left[\sum_{k}a_{k}\hat{a}_{k}^{\dagger}\right]^{Q}\left|0_{\textrm{ex}}\right\rangle 
\langle \psi_{\textrm{ph}}(t)|\psi_{\textrm{ph}}(t)\rangle \nonumber \\
 & =\frac{1}{Q!}\left\langle 0_{\textrm{ex}}\right|
\sum_{q_{1}+\ldots+q_{n}=Q}
{\scriptstyle 
Q!\frac{\left(a_{1}^{*}\hat{a}_{1}\right)^{q_{1}}}{q_{1}!}\frac{\left(a_{2}^{*}\hat{a}_{2}\right)^{q_{2}}}{q_{2}!}\ldots\frac{\left(a_{n}^{*}\hat{a}_{n}\right)^{q_{n}}}{q_{n}!}\hat{a}_{1}^{\dagger}\hat{a}_{2}
}
\sum_{k_{1}+\ldots+k_{n}=Q}
{\scriptstyle 
Q!\frac{\left(a_{1}\hat{a}_{1}^{\dagger}\right)^{k_{1}}}{k_{1}!}\frac{\left(a_{2}\hat{a}_{2}^{\dagger}\right)^{k_{2}}}{k_{2}!}\ldots\frac{\left(a_{n}\hat{a}_{n}^{\dagger}\right)^{k_{n}}}{k_{n}!}
}
\left|0_{\textrm{ex}}\right\rangle \nonumber \\
 & =Q!\left\langle 0_{\textrm{ex}}\right|\sum_{k_{1}+\ldots+k_{n}=Q}
{\scriptstyle
\frac{\left(a_{1}^{*}\hat{a}_{1}\right)^{k_{1}+1}}{(k_{1}+1)!}\hat{a}_{1}^{\dagger}\frac{\left(a_{1}\hat{a}_{1}^{\dagger}\right)^{k_{1}}}{k_{1}!}\frac{\left(a_{2}^{*}\hat{a}_{2}\right)^{k_{2}-1}}{(k_{2}-1)!}\hat{a}_{2}\frac{\left(a_{2}\hat{a}_{2}^{\dagger}\right)^{k_{2}}}{k_{2}!}\ldots\frac{\left(a_{n}^{*}\hat{a}_{n}\right)^{k_{n}}}{k_{n}!}\frac{\left(a_{n}\hat{a}_{n}^{\dagger}\right)^{k_{n}}}{k_{n}!}
}
\left|0_{\textrm{ex}}\right\rangle \nonumber \\
 & =Q!\left\langle 0_{\textrm{ex}}\right|\sum_{k_{1}+\ldots+k_{n}=Q}
{\scriptstyle
a_{1}^{*}\frac{\left(a_{1}^{*}a_{1}\right)^{k_{1}}}{k_{1}!}\frac{\left(\hat{a}_{1}\right)^{k_{1}+1}\left(\hat{a}_{1}^{\dagger}\right)^{k_{1}+1}}{(k_{1}+1)!}a_{2}\frac{\left(a_{2}^{*}a_{2}\right)^{k_{2}-1}}{(k_{2}-1)!}\frac{\left(\hat{a}_{2}\right)^{k_{2}}\left(\hat{a}_{2}^{\dagger}\right)^{k_{2}}}{k_{2}!}\ldots\frac{\left(a_{n}^{*}a_{n}\right)^{k_{n}}}{k_{n}!}\frac{\left(\hat{a}_{n}\right)^{k_{n}}\left(\hat{a}_{n}^{\dagger}\right)^{k_{n}}}{k_{n}!}
}
\left|0_{\textrm{ex}}\right\rangle \nonumber \\
 & =Q!a_{1}^{*}a_{2}\sum_{k_{1}+\ldots+k_{n}=Q}\frac{\left(a_{1}^{*}a_{1}\right)^{k_{1}}}{k_{1}!}\frac{\left(a_{2}^{*}a_{2}\right)^{k_{2}-1}}{(k_{2}-1)!}\ldots\frac{\left(a_{n}^{*}a_{n}\right)^{k_{n}}}{k_{n}!}\nonumber \\
 & =Qa_{1}^{*}a_{2}(Q-1)!\sum_{m_{1}+m_{2}+\ldots+m_{n}=Q-1}\frac{\left(a_{1}^{*}a_{1}\right)^{m_{1}}}{m_{1}!}\frac{\left(a_{2}^{*}a_{2}\right)^{m_{2}}}{m_{2}!}\ldots\frac{\left(a_{n}^{*}a_{n}\right)^{m_{n}}}{m_{n}!}\nonumber \\
 & =Qa_{1}^{*}a_{2}\left(\sum_{m}a_{m}^{*}a_{m}\right)^{Q-1}=Qa_{1}^{*}a_{2}\times1^{Q-1}=Qa_{1}^{*}a_{2}
\end{align}
In the above calculation, we have dropped the repeated index among
$n,\alpha$ in order to ease the notation. Similarly, one obtains
$\langle\hat{a}_{n}^{\dagger}\hat{a}_{n\pm1}\rangle=Qa_{n}^{*}a_{n\pm1}$
and $\langle\hat{a}_{n}^{\dagger}\hat{a}_{n}\rangle=Qa_{n}^{*}a_{n}$
\cite{Georgiev2020a}.

From the normalization of quantum probability
$\sum_{n,\alpha}\left|a_{n,\alpha}\right|^{2}=1$, it follows that
the expectation value for the exciton number operator summed over
all sites is also conserved 
\begin{equation}
\left\langle\sum_{n,\alpha}\hat{a}_{n,\alpha}^{\dagger}\hat{a}_{n,\alpha}\right\rangle=\sum_{n,\alpha}Qa_{n,\alpha}^{*}a_{n,\alpha}\times1^{Q-1}=Q\sum_{n,\alpha}\left|a_{n,\alpha}\right|^{2}=Q\times1=Q
\end{equation}
Therefore, all simulation plots could be also interpreted as showing
the expectation value for the exciton number operator provided that
all values displayed in the scale bars are multiplied by $Q=3$.

\section{Supplementary material}


\renewcommand{\thefigure}{S\arabic{figure}}
\setcounter{figure}{0}

We provide Supplementary Figs.~S1--S6, for each of the main Figs.~1--6, with the corresponding sums of exciton quantum probabilities on the right side of barrier, inside the barrier, or on the left side of the barrier.
We also indicate with arrows the time points that have maximal quantum probability for finding each of the three excitons ($Q=3$) on the left side of the barrier.

We also present Supplementary Videos 1--3 with captions, for each of the main Figs.~7a, 7b and 1b.

\begin{figure}
\begin{centering}
\includegraphics[width=160mm]{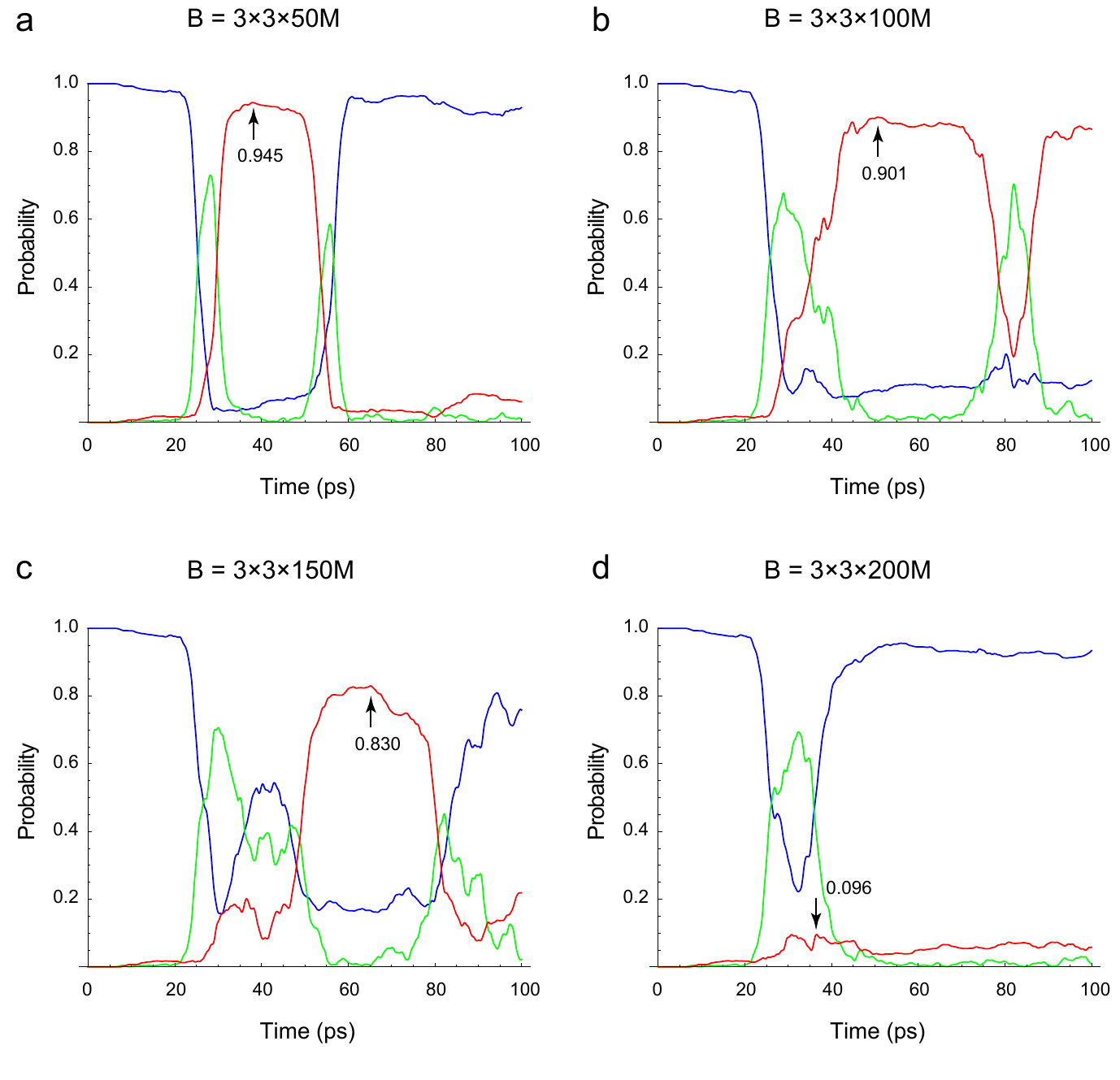}
\par\end{centering}

\caption{Dynamics of the exciton quantum probabilities on the right side of barrier
$\displaystyle \sum_{n=1}^{25}\sum_{\alpha=0}^{2}\left|a_{n,\alpha}\right|^{2}$ (blue line),
inside the barrier $\displaystyle \sum_{n=26}^{28}\sum_{\alpha=0}^{2}\left|a_{n,\alpha}\right|^{2}$ (green line), 
or on the left side of the barrier $\displaystyle \sum_{n=29}^{40}\sum_{\alpha=0}^{2}\left|a_{n,\alpha}\right|^{2}$ (red line)
in the simulation of a molecular soliton (Fig.~1 in main text) initiated by a
sech-squared exciton pulse with $Q=3$ amide~I exciton quanta spread initially over 5 peptide groups
that undergoes tunneling through a concentric massive barrier $B$
centered on all three spines ($3\times$) extending over three peptide
groups ($3\times$), $n$=26--28, each of which with increased effective
mass up to $50M$ (a), $100M$ (b), $150M$ (c) or $200M$ (d).
Arrows indicate the time points that have the maximal quantum probability for finding each of the three excitons ($Q=3$) on the left side of the barrier.}
\end{figure}

\begin{figure}
\begin{centering}
\includegraphics[width=160mm]{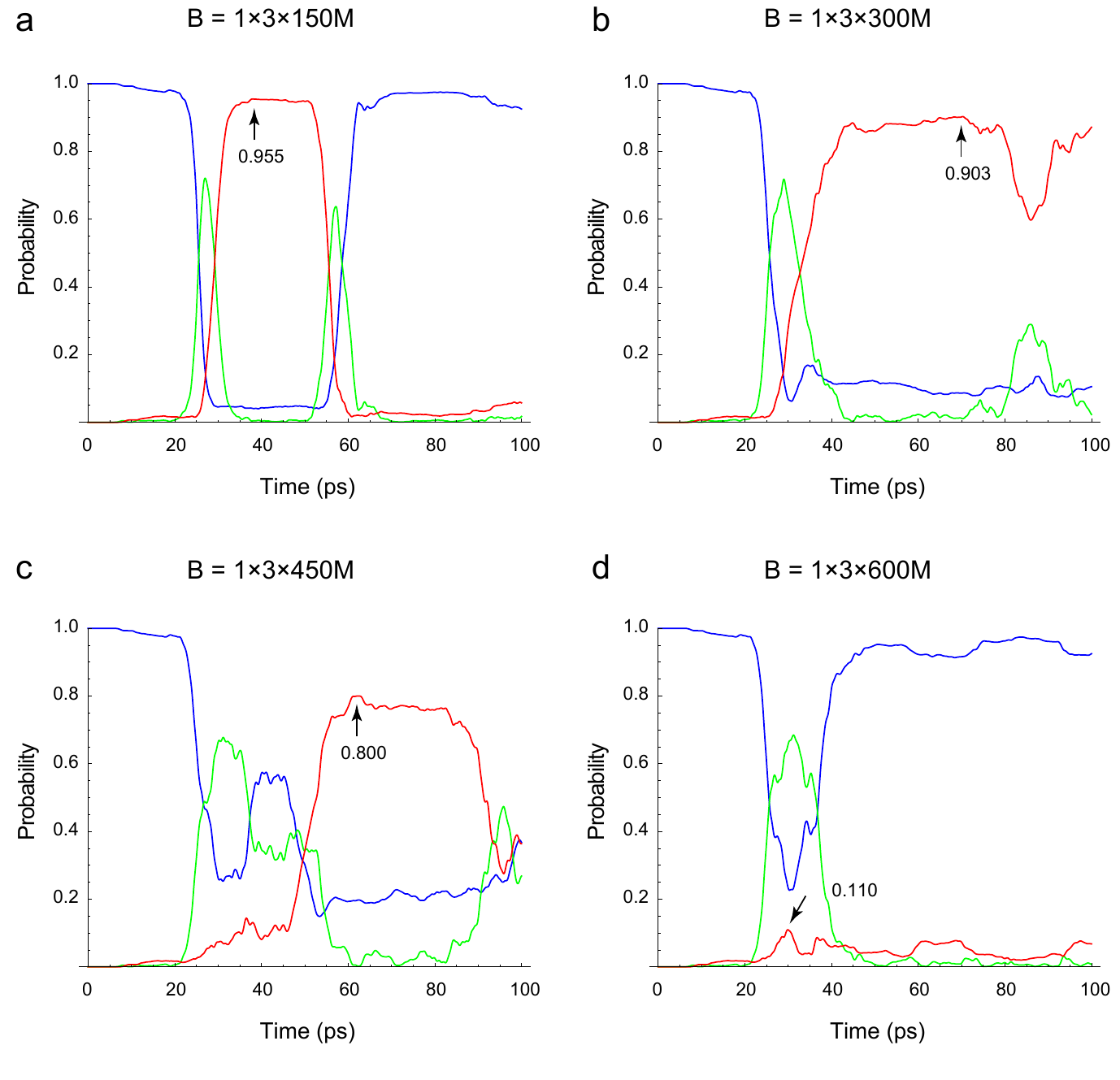}
\par\end{centering}

\caption{Dynamics of the exciton quantum probabilities on the right side of barrier
$\displaystyle \sum_{n=1}^{25}\sum_{\alpha=0}^{2}\left|a_{n,\alpha}\right|^{2}$ (blue line),
inside the barrier $\displaystyle \sum_{n=26}^{28}\sum_{\alpha=0}^{2}\left|a_{n,\alpha}\right|^{2}$ (green line), 
or on the left side of the barrier $\displaystyle \sum_{n=29}^{40}\sum_{\alpha=0}^{2}\left|a_{n,\alpha}\right|^{2}$ (red line)
in the simulation of a molecular soliton (Fig.~2 in main text) initiated by a
sech-squared exciton pulse with $Q=3$ amide~I exciton quanta spread initially over 5 peptide groups
that undergoes tunneling through an excentric massive barrier $B$
placed on a single spine ($1\times$) extending along three peptide
groups ($3\times$), $n$=26--28, each of which with increased effective
mass up to $150M$ (a), $300M$ (b), $450M$ (c) or $600M$ (d).
Arrows indicate the time points that have the maximal quantum probability for finding each of the three excitons ($Q=3$) on the left side of the barrier.}
\end{figure}

\begin{figure}
\begin{centering}
\includegraphics[width=160mm]{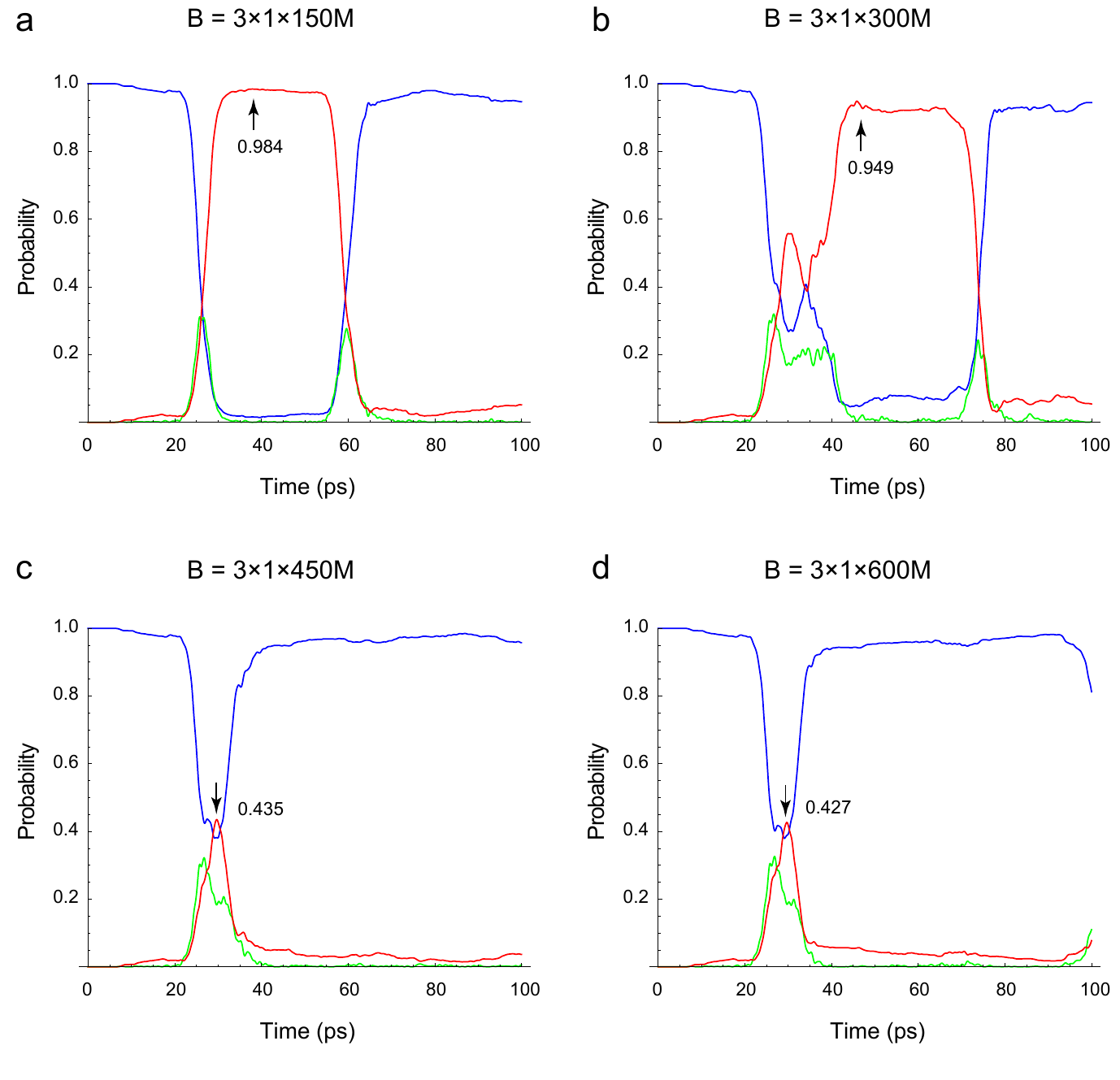}
\par\end{centering}

\caption{Dynamics of the exciton quantum probabilities on the right side of barrier
$\displaystyle \sum_{n=1}^{25}\sum_{\alpha=0}^{2}\left|a_{n,\alpha}\right|^{2}$ (blue line),
inside the barrier $\displaystyle \sum_{n=26}^{26}\sum_{\alpha=0}^{2}\left|a_{n,\alpha}\right|^{2}$ (green line), 
or on the left side of the barrier $\displaystyle \sum_{n=27}^{40}\sum_{\alpha=0}^{2}\left|a_{n,\alpha}\right|^{2}$ (red line)
in the simulation of a molecular soliton (Fig.~3 in main text) initiated by a
sech-squared exciton pulse with $Q=3$ amide~I exciton quanta spread initially over 5 peptide groups
that undergoes tunneling through a concentric massive barrier $B$
centered on all three spines ($3\times$) extending along a single
peptide group ($1\times$), $n=26$, with increased effective mass
up to $150M$ (a), $300M$ (b), $450M$ (c) or $600M$ (d).
Arrows indicate the time points that have the maximal quantum probability for finding each of the three excitons ($Q=3$) on the left side of the barrier.}
\end{figure}

\begin{figure}
\begin{centering}
\includegraphics[width=160mm]{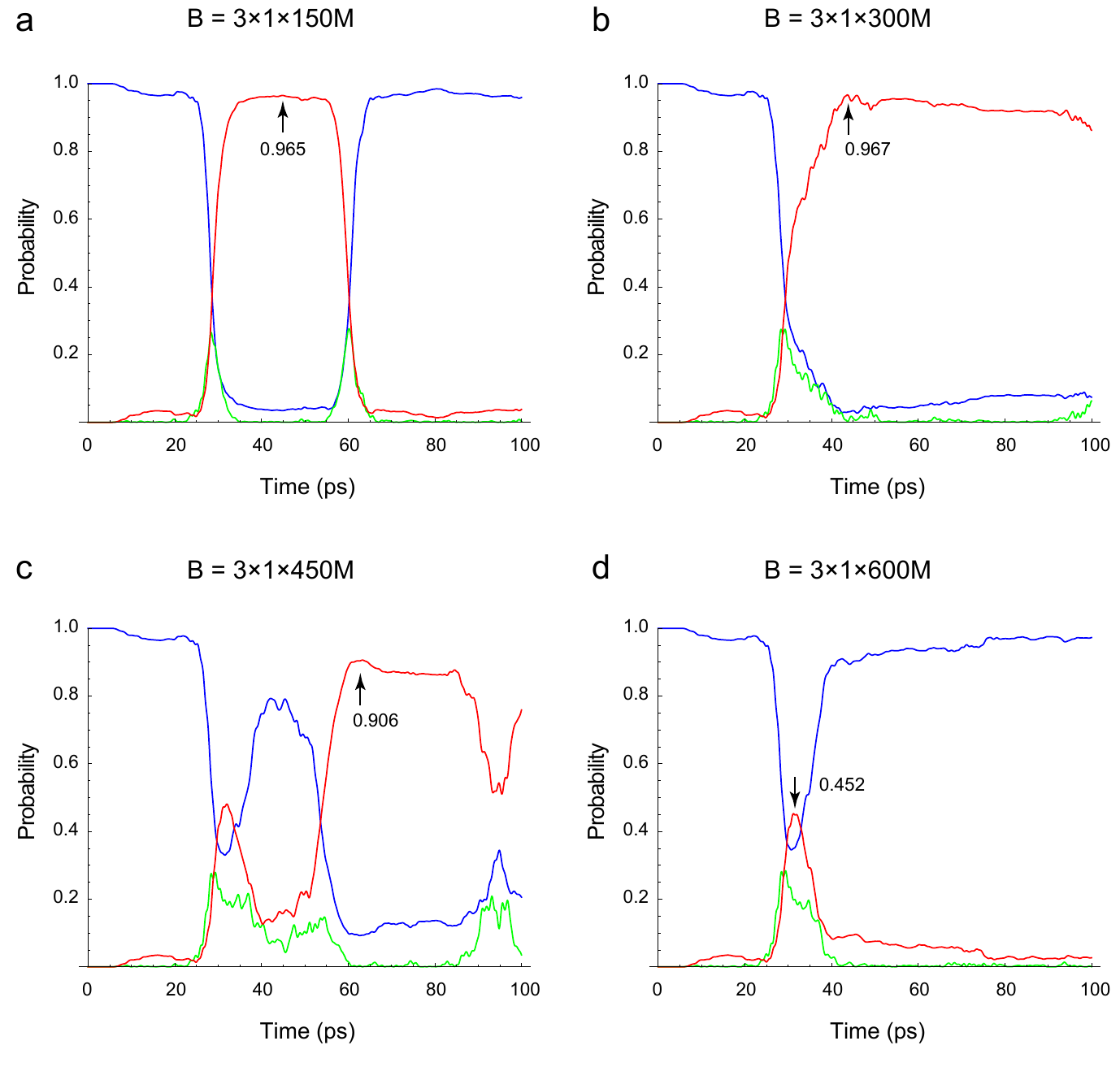}
\par\end{centering}

\caption{Dynamics of the exciton quantum probabilities on the right side of barrier
$\displaystyle \sum_{n=1}^{25}\sum_{\alpha=0}^{2}\left|a_{n,\alpha}\right|^{2}$ (blue line),
inside the barrier $\displaystyle \sum_{n=26}^{26}\sum_{\alpha=0}^{2}\left|a_{n,\alpha}\right|^{2}$ (green line), 
or on the left side of the barrier $\displaystyle \sum_{n=27}^{40}\sum_{\alpha=0}^{2}\left|a_{n,\alpha}\right|^{2}$ (red line)
in the simulation of a molecular soliton (Fig.~4 in main text) initiated by a
sech-squared exciton pulse with $Q=3$ amide~I exciton quanta spread initially over 7 peptide groups
that undergoes tunneling through a concentric massive barrier $B$
centered on all three spines ($3\times$) extending along a single
peptide group ($1\times$), $n=26$, with increased effective mass
up to $150M$ (a), $300M$ (b), $450M$ (c) or $600M$ (d).
Arrows indicate the time points that have the maximal quantum probability for finding each of the three excitons ($Q=3$) on the left side of the barrier.}
\end{figure}

\begin{figure}
\begin{centering}
\includegraphics[width=160mm]{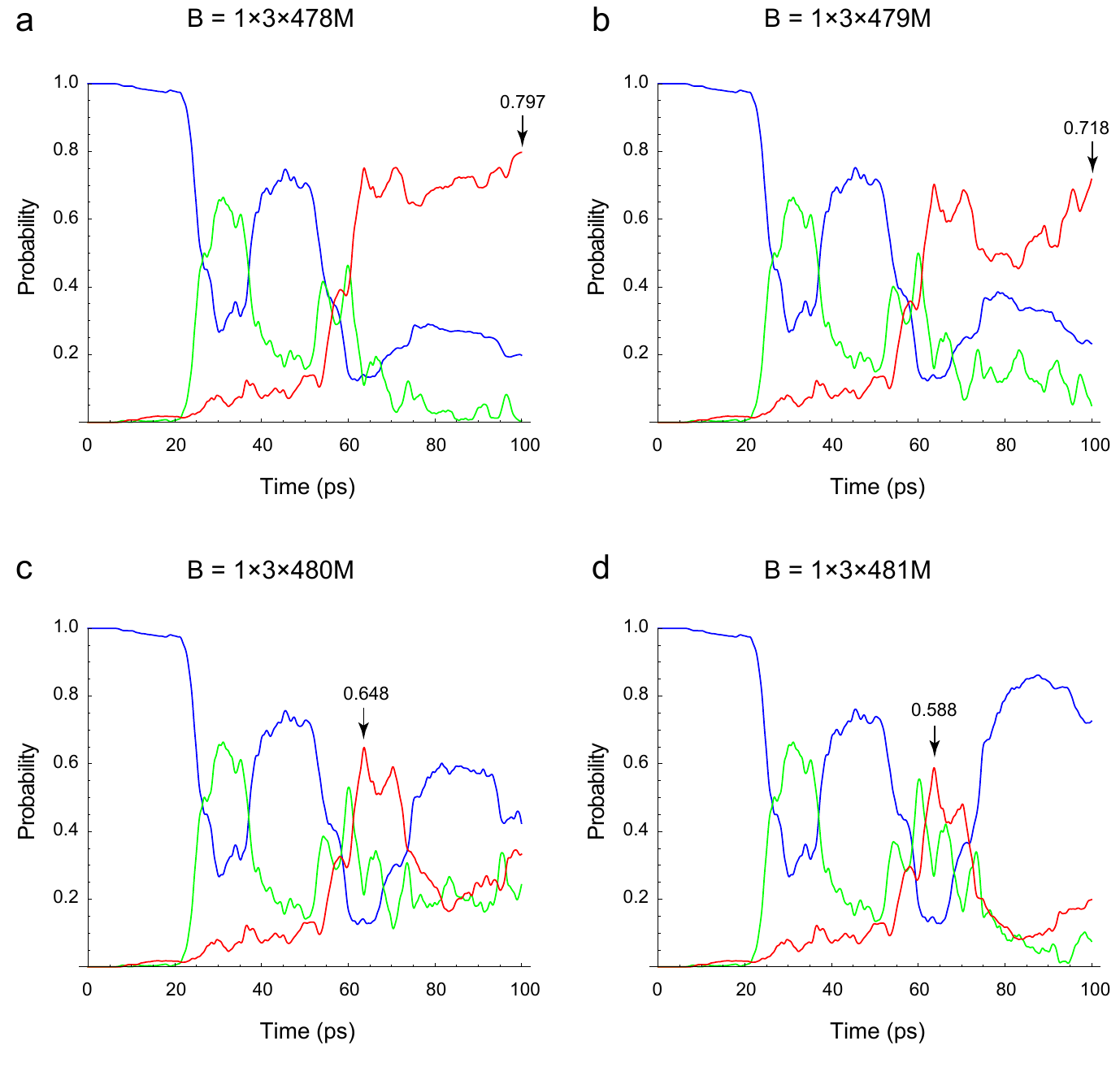}
\par\end{centering}

\caption{Dynamics of the exciton quantum probabilities on the right side of barrier
$\displaystyle \sum_{n=1}^{25}\sum_{\alpha=0}^{2}\left|a_{n,\alpha}\right|^{2}$ (blue line),
inside the barrier $\displaystyle \sum_{n=26}^{28}\sum_{\alpha=0}^{2}\left|a_{n,\alpha}\right|^{2}$ (green line), 
or on the left side of the barrier $\displaystyle \sum_{n=29}^{40}\sum_{\alpha=0}^{2}\left|a_{n,\alpha}\right|^{2}$ (red line)
in the simulation of a molecular soliton (Fig.~5 in main text) initiated by a sech-squared
exciton pulse with $Q=3$ amide~I exciton quanta spread initially over 5 peptide groups that impacts
upon an excentric massive barrier $B$ placed on a single spine ($1\times$)
extending along three peptide groups ($3\times$), $n$=26--28, each
of which with increased effective mass up to $478M$ (a), $479M$
(b), $480M$ (c) or $481M$ (d).
Arrows indicate the time points that have the maximal quantum probability for finding each of the three excitons ($Q=3$) on the left side of the barrier.}
\end{figure}

\begin{figure}
\begin{centering}
\includegraphics[width=160mm]{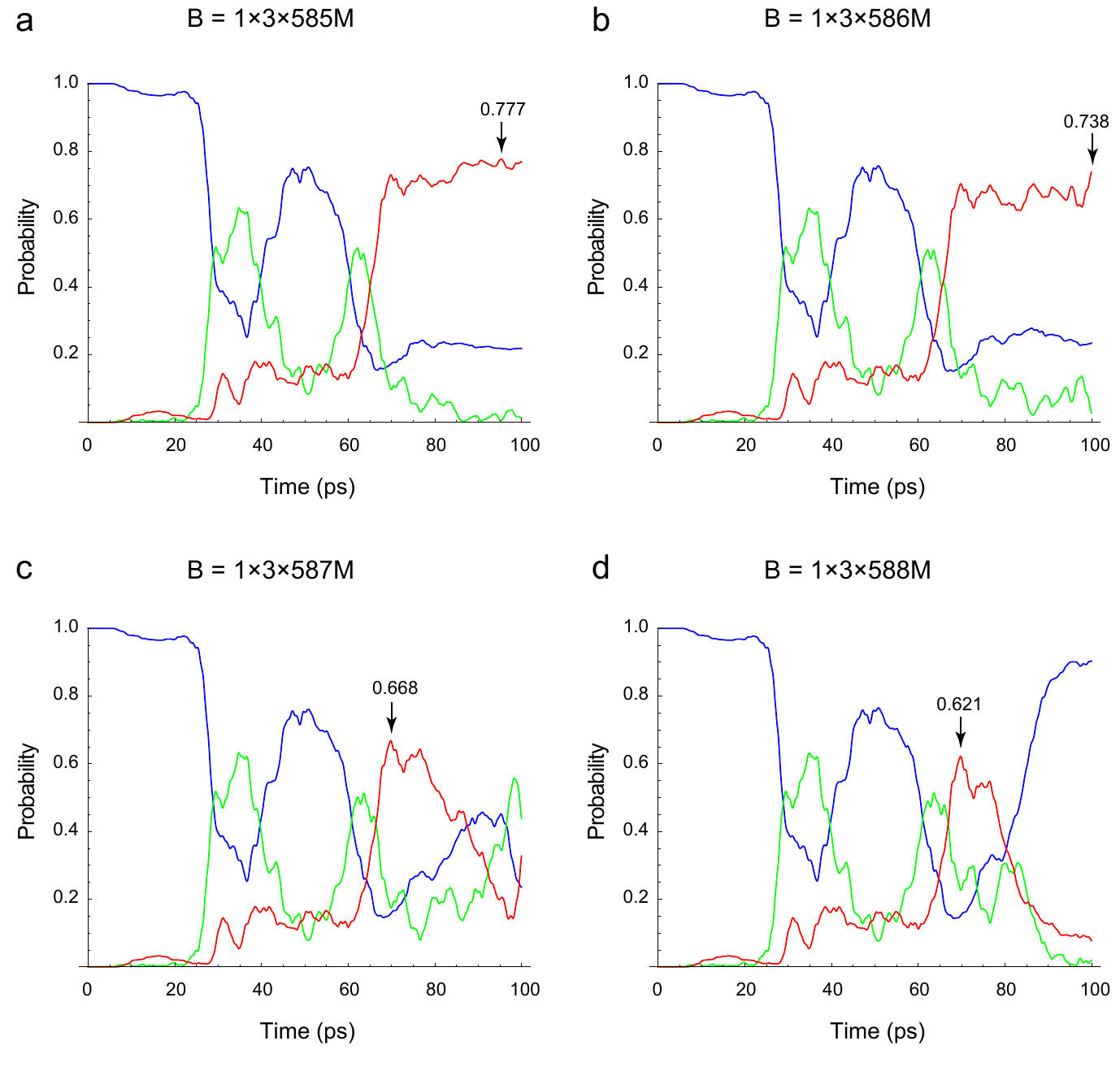}
\par\end{centering}

\caption{Dynamics of the exciton quantum probabilities on the right side of barrier
$\displaystyle \sum_{n=1}^{25}\sum_{\alpha=0}^{2}\left|a_{n,\alpha}\right|^{2}$ (blue line),
inside the barrier $\displaystyle \sum_{n=26}^{28}\sum_{\alpha=0}^{2}\left|a_{n,\alpha}\right|^{2}$ (green line), 
or on the left side of the barrier $\displaystyle \sum_{n=29}^{40}\sum_{\alpha=0}^{2}\left|a_{n,\alpha}\right|^{2}$ (red line)
in the simulation of a molecular soliton (Fig.~6 in main text) initiated by a sech-squared
exciton pulse with $Q=3$ amide~I exciton quanta spread initially over 7 peptide groups that impacts
upon an excentric massive barrier $B$ placed on a single spine ($1\times$)
extending along three peptide groups ($3\times$), $n$=26--28, each
of which with increased effective mass up to $585M$ (a), $586M$
(b), $587M$ (c) or $588M$ (d). 
Arrows indicate the time points that have the maximal quantum probability for finding each of the three excitons ($Q=3$) on the left side of the barrier.}
\end{figure}

\pagebreak

\section*{Supplementary Videos}

\noindent Supplementary Videos 1--3
can be found online at \url{https://doi.org/10.1016/j.physleta.2022.128319}

\paragraph{Supplementary Video 1} The quantum dynamics of the real and imaginary parts, $\textrm{Re}(a_{n,\alpha})$ and $\textrm{Im}(a_{n,\alpha})$, of the exciton quantum probability amplitudes $a_{n,\alpha}$ for each of the three spines $\alpha\in{0,1,2}$ of a molecular soliton initiated by a
sech-squared exciton pulse with a single amide~I quantum $Q=1$, spread initially over $n=5$ peptide groups
that undergoes tunneling through a concentric massive barrier $B$ centered on all three spines~($3\times$) extending over three peptide
groups~($3\times$), $n$=26--28, each of which with increased effective mass up to $100M$.
The massive barrier, whose location is indicated with two horizontal black lines, is able to disperse the soliton by the end of the simulation period of 100~ps.
Contour plot of the exciton quantum probabilities for this simulation is shown in main Fig.~7a.

\paragraph{Supplementary Video 2} The quantum dynamics of the real and imaginary parts, $\textrm{Re}(a_{n,\alpha})$ and $\textrm{Im}(a_{n,\alpha})$, of the exciton quantum probability amplitudes $a_{n,\alpha}$ for each of the three spines $\alpha\in{0,1,2}$ of a molecular soliton initiated by a
sech-squared exciton pulse with two amide~I quanta $Q=2$, spread initially over $n=5$ peptide groups
that undergoes tunneling through a concentric massive barrier $B$ centered on all three spines~($3\times$) extending over three peptide
groups~($3\times$), $n$=26--28, each of which with increased effective mass up to $100M$.
The massive barrier, whose location is indicated with two horizontal black lines, is able to disperse the soliton by the end of the simulation period of 100~ps.
Contour plot of the exciton quantum probabilities for this simulation is shown in main Fig.~7b.

\paragraph{Supplementary Video 3} The quantum dynamics of the real and imaginary parts, $\textrm{Re}(a_{n,\alpha})$ and $\textrm{Im}(a_{n,\alpha})$, of the exciton quantum probability amplitudes $a_{n,\alpha}$ for each of the three spines $\alpha\in{0,1,2}$ of a molecular soliton initiated by a
sech-squared exciton pulse with three amide~I quanta $Q=3$, spread initially over $n=5$ peptide groups
that undergoes tunneling through a concentric massive barrier $B$ centered on all three spines~($3\times$) extending over three peptide
groups~($3\times$), $n$=26--28, each of which with increased effective mass up to $100M$.
The massive barrier, whose location is indicated with two horizontal black lines, is unable to disperse the soliton by the end of the simulation period of 100~ps.
Contour plot of the exciton quantum probabilities for this simulation is shown in main Fig.~1b.

\pagebreak


\begin{thebibliography}{10}
\expandafter\ifx\csname url\endcsname\relax
  \def\url#1{\texttt{#1}}\fi
\expandafter\ifx\csname urlprefix\endcsname\relax\def\urlprefix{URL }\fi
\expandafter\ifx\csname href\endcsname\relax
  \def\href#1#2{#2} \def\path#1{#1}\fi

\bibitem{Rodwell2018}
V.~W. Rodwell, D.~Bender, K.~M. Botham, P.~J. Kennelly, P.~A. Weil, Harper's
  Illustrated Biochemistry, 31st Edition, McGraw-Hill Education, New York,
  2018.

\bibitem{Chiesa2020}
G.~Chiesa, S.~Kiriakov, A.~S. Khalil, Protein assembly systems in natural and
  synthetic biology, BMC Biology 18~(1) (2020) 35.
\newblock \href {https://doi.org/10.1186/s12915-020-0751-4}
  {\path{doi:10.1186/s12915-020-0751-4}}.

\bibitem{Stollar2020}
E.~J. Stollar, D.~P. Smith, Uncovering protein structure, Essays in
  Biochemistry 64~(4) (2020) 649--680.
\newblock \href {https://doi.org/10.1042/ebc20190042}
  {\path{doi:10.1042/ebc20190042}}.

\bibitem{ODonoghue2010}
S.~I. O'Donoghue, D.~S. Goodsell, A.~S. Frangakis, F.~Jossinet, R.~A.
  Laskowski, M.~Nilges, H.~R. Saibil, A.~Schafferhans, R.~C. Wade, E.~Westhof,
  A.~J. Olson, Visualization of macromolecular structures, Nature Methods 7~(3)
  (2010) S42--S55.
\newblock \href {https://doi.org/10.1038/nmeth.1427}
  {\path{doi:10.1038/nmeth.1427}}.

\bibitem{Durand2013}
A.~Durand, G.~Papai, P.~Schultz, Structure, assembly and dynamics of
  macromolecular complexes by single particle cryo-electron microscopy, Journal
  of Nanobiotechnology 11~(1) (2013) S4.
\newblock \href {https://doi.org/10.1186/1477-3155-11-S1-S4}
  {\path{doi:10.1186/1477-3155-11-S1-S4}}.

\bibitem{Larochelle2015}
S.~Larochelle, Putting the pieces together, Nature Methods 12~(1) (2015) 21.
\newblock \href {https://doi.org/10.1038/nmeth.3530}
  {\path{doi:10.1038/nmeth.3530}}.

\bibitem{Creighton1993}
T.~E. Creighton, Proteins: Structures and Molecular Properties, W. H. Freeman,
  New York, 1993.

\bibitem{Kessel2018}
A.~Kessel, N.~Ben-Tal, Introduction to Proteins: Structure, Function, and
  Motion, 2nd Edition, Chapman and Hall/CRC, New York, 2018.
\newblock \href {https://doi.org/10.1201/9781315113876}
  {\path{doi:10.1201/9781315113876}}.

\bibitem{Georgiev2020c}
D.~D. Georgiev, S.~K. Kolev, E.~Cohen, J.~F. Glazebrook, Computational capacity
  of pyramidal neurons in the cerebral cortex, Brain Research 1748 (2020)
  147069.
\newblock \href {https://doi.org/10.1016/j.brainres.2020.147069}
  {\path{doi:10.1016/j.brainres.2020.147069}}.

\bibitem{Pauling1951}
L.~Pauling, R.~B. Corey, H.~R. Branson, The structure of proteins: two
  hydrogen-bonded helical configurations of the polypeptide chain, Proceedings
  of the National Academy of Sciences of the United States of America 37~(4)
  (1951) 205--211.
\newblock \href {https://doi.org/10.1073/pnas.37.4.205}
  {\path{doi:10.1073/pnas.37.4.205}}.

\bibitem{Sivaramakrishnan2008}
S.~Sivaramakrishnan, B.~J. Spink, A.~Y.~L. Sim, S.~Doniach, J.~A. Spudich,
  Dynamic charge interactions create surprising rigidity in the {ER/K}
  $\alpha$-helical protein motif, Proceedings of the National Academy of
  Sciences 105~(36) (2008) 13356--13361.
\newblock \href {https://doi.org/10.1073/pnas.0806256105}
  {\path{doi:10.1073/pnas.0806256105}}.

\bibitem{Davydov1976}
A.~S. Davydov, N.~I. Kislukha, Solitons in one-dimensional molecular chains,
  Physica Status Solidi B 75~(2) (1976) 735--742.
\newblock \href {https://doi.org/10.1002/pssb.2220750238}
  {\path{doi:10.1002/pssb.2220750238}}.

\bibitem{Davydov1979}
A.~S. Davydov, Solitons in molecular systems, Physica Scripta 20~(3-4) (1979)
  387--394.
\newblock \href {https://doi.org/10.1088/0031-8949/20/3-4/013}
  {\path{doi:10.1088/0031-8949/20/3-4/013}}.

\bibitem{Davydov1979b}
A.~S. Davydov, Solitons, bioenergetics, and the mechanism of muscle
  contraction, International Journal of Quantum Chemistry 16~(1) (1979) 5--17.
\newblock \href {https://doi.org/10.1002/qua.560160104}
  {\path{doi:10.1002/qua.560160104}}.

\bibitem{Davydov1981}
A.~S. Davydov, The role of solitons in the energy and electron transfer in
  one-dimensional molecular systems, Physica D: Nonlinear Phenomena 3~(1-2)
  (1981) 1--22.
\newblock \href {https://doi.org/10.1016/0167-2789(81)90116-0}
  {\path{doi:10.1016/0167-2789(81)90116-0}}.

\bibitem{Davydov1982}
A.~S. Davydov, Solitons in quasi-one-dimensional molecular structures, Soviet
  Physics Uspekhi 25~(12) (1982) 898--918.
\newblock \href {https://doi.org/10.1070/pu1982v025n12abeh005012}
  {\path{doi:10.1070/pu1982v025n12abeh005012}}.

\bibitem{Davydov1986}
A.~S. Davydov, Quantum theory of the motion of a quasi-particle in a molecular
  chain with thermal vibrations taken into account, Physica Status Solidi B
  138~(2) (1986) 559--576.
\newblock \href {https://doi.org/10.1002/pssb.2221380221}
  {\path{doi:10.1002/pssb.2221380221}}.

\bibitem{Brizhik1983}
L.~S. Brizhik, A.~S. Davydov, Soliton excitations in one-dimensional molecular
  systems, Physica Status Solidi B 115~(2) (1983) 615--630.
\newblock \href {https://doi.org/10.1002/pssb.2221150233}
  {\path{doi:10.1002/pssb.2221150233}}.

\bibitem{Brizhik1988}
L.~S. Brizhik, Y.~B. Gaididei, A.~A. Vakhnenko, V.~A. Vakhnenko, Soliton
  generation in semi-infinite molecular chains, Physica Status Solidi B 146~(2)
  (1988) 605--612.
\newblock \href {https://doi.org/10.1002/pssb.2221460221}
  {\path{doi:10.1002/pssb.2221460221}}.

\bibitem{Brizhik1993}
L.~S. Brizhik, Soliton generation in molecular chains, Physical Review B 48~(5)
  (1993) 3142--3144.
\newblock \href {https://doi.org/10.1103/PhysRevB.48.3142}
  {\path{doi:10.1103/PhysRevB.48.3142}}.

\bibitem{Kivshar1989}
Y.~S. Kivshar, B.~A. Malomed, Dynamics of solitons in nearly integrable
  systems, Reviews of Modern Physics 61~(4) (1989) 763--915.
\newblock \href {https://doi.org/10.1103/RevModPhys.61.763}
  {\path{doi:10.1103/RevModPhys.61.763}}.

\bibitem{Brizhik2004}
L.~S. Brizhik, A.~A. Eremko, B.~Piette, W.~J.~M. Zakrzewski, Solitons in
  $\alpha$-helical proteins, Physical Review E 70~(3) (2004) 031914.
\newblock \href {https://doi.org/10.1103/PhysRevE.70.031914}
  {\path{doi:10.1103/PhysRevE.70.031914}}.

\bibitem{Brizhik2006}
L.~Brizhik, A.~Eremko, B.~Piette, W.~Zakrzewski, Charge and energy transfer by
  solitons in low-dimensional nanosystems with helical structure, Chemical
  Physics 324~(1) (2006) 259--266.
\newblock \href {https://doi.org/10.1016/j.chemphys.2006.01.033}
  {\path{doi:10.1016/j.chemphys.2006.01.033}}.

\bibitem{Brizhik2019}
L.~S. Brizhik, J.~Luo, B.~M. A.~G. Piette, W.~J. Zakrzewski, Long-range
  donor-acceptor electron transport mediated by $\alpha$-helices, Physical
  Review E 100~(6) (2019) 062205.
\newblock \href {https://doi.org/10.1103/PhysRevE.100.062205}
  {\path{doi:10.1103/PhysRevE.100.062205}}.

\bibitem{Georgiev2019a}
D.~D. Georgiev, J.~F. Glazebrook, On the quantum dynamics of {D}avydov solitons
  in protein $\alpha$-helices, Physica A: Statistical Mechanics and its
  Applications 517 (2019) 257--269.
\newblock \href {https://doi.org/10.1016/j.physa.2018.11.026}
  {\path{doi:10.1016/j.physa.2018.11.026}}.

\bibitem{Georgiev2020a}
D.~D. Georgiev, J.~F. Glazebrook, Quantum transport and utilization of free
  energy in protein $\alpha$-helices, Advances in Quantum Chemistry 82 (2020)
  253--300.
\newblock \href {https://doi.org/10.1016/bs.aiq.2020.02.001}
  {\path{doi:10.1016/bs.aiq.2020.02.001}}.

\bibitem{Georgiev2022}
D.~D. Georgiev, J.~F. Glazebrook, Thermal stability of solitons in protein
  $\alpha$-helices, Chaos, Solitons and Fractals 155 (2022) 111644.
\newblock \href {https://doi.org/10.1016/j.chaos.2021.111644}
  {\path{doi:10.1016/j.chaos.2021.111644}}.

\bibitem{Davydov1978}
A.~S. Davydov, A.~A. Eremko, A.~I. Sergienko, Solitons in $\alpha$-helical
  protein molecules, Ukrainskii Fizichnii Zhurnal 23~(6) (1978) 983--993.

\bibitem{Hyman1981}
J.~M. Hyman, D.~W. McLaughlin, A.~C. Scott, On {D}avydov's alpha-helix
  solitons, Physica D: Nonlinear Phenomena 3~(1-2) (1981) 23--44.
\newblock \href {https://doi.org/10.1016/0167-2789(81)90117-2}
  {\path{doi:10.1016/0167-2789(81)90117-2}}.

\bibitem{Luo2017}
J.~Luo, B.~M. A.~G. Piette, A generalised {D}avydov--{S}cott model for polarons
  in linear peptide chains, European Physical Journal B 90~(8) (2017) 155.
\newblock \href {https://doi.org/10.1140/epjb/e2017-80209-2}
  {\path{doi:10.1140/epjb/e2017-80209-2}}.

\bibitem{Nevskaya1976}
N.~A. Nevskaya, Y.~N. Chirgadze, Infrared spectra and resonance interactions of
  amide-{I} and {II} vibrations of $\alpha$-helix, Biopolymers 15~(4) (1976)
  637--648.
\newblock \href {https://doi.org/10.1002/bip.1976.360150404}
  {\path{doi:10.1002/bip.1976.360150404}}.

\bibitem{Itoh1972}
K.~Itoh, T.~Shimanouchi, Vibrational spectra of crystalline formamide, Journal
  of Molecular Spectroscopy 42~(1) (1972) 86--99.
\newblock \href {https://doi.org/10.1016/0022-2852(72)90146-4}
  {\path{doi:10.1016/0022-2852(72)90146-4}}.

\bibitem{Savin1993}
A.~V. Savin, A.~V. Zolotaryuk, Dynamics of the amide-{I} excitation in a
  molecular chain with thermalized acoustic and optical modes, Physica D:
  Nonlinear Phenomena 68~(1) (1993) 59--64.
\newblock \href {https://doi.org/10.1016/0167-2789(93)90029-Z}
  {\path{doi:10.1016/0167-2789(93)90029-Z}}.

\bibitem{Scott1992}
A.~C. Scott, Davydov's soliton, Physics Reports 217~(1) (1992) 1--67.
\newblock \href {https://doi.org/10.1016/0370-1573(92)90093-F}
  {\path{doi:10.1016/0370-1573(92)90093-F}}.

\bibitem{Georgiev2019b}
D.~D. Georgiev, J.~F. Glazebrook, Quantum tunneling of {D}avydov solitons
  through massive barriers, Chaos, Solitons and Fractals 123 (2019) 275--293.
\newblock \href {https://doi.org/10.1016/j.chaos.2019.04.013}
  {\path{doi:10.1016/j.chaos.2019.04.013}}.

\bibitem{Brizhik2003}
L.~S. Brizhik, A.~A. Eremko, B.~M. A.~G. Piette, W.~J. Zakrzewski, Spontaneous
  localization of electrons in two-dimensional lattices within the adiabatic
  approximation, Journal of Mathematical Physics 44~(9) (2003) 3689--3697.
\newblock \href {https://doi.org/10.1063/1.1592873}
  {\path{doi:10.1063/1.1592873}}.

\bibitem{Pitaevskii2003}
L.~Pitaevskii, S.~Stringari, Bose--Einstein Condensation, Oxford University
  Press, Oxford, 2003.

\bibitem{Zolotaryuk1988}
A.~V. Zolotaryuk, Many-particle {D}avydov solitons, Physics of Many Particle
  Systems 13 (1988) 40--52.

\bibitem{Glauber1963a}
R.~J. Glauber, The quantum theory of optical coherence, Physical Review 130~(6)
  (1963) 2529--2539.
\newblock \href {https://doi.org/10.1103/PhysRev.130.2529}
  {\path{doi:10.1103/PhysRev.130.2529}}.

\bibitem{Glauber1963b}
R.~J. Glauber, Coherent and incoherent states of the radiation field, Physical
  Review 131~(6) (1963) 2766--2788.
\newblock \href {https://doi.org/10.1103/PhysRev.131.2766}
  {\path{doi:10.1103/PhysRev.131.2766}}.

\bibitem{Kerr1987}
W.~C. Kerr, P.~S. Lomdahl, Quantum-mechanical derivation of the equations of
  motion for {D}avydov solitons, Physical Review B 35~(7) (1987) 3629--3632.
\newblock \href {https://doi.org/10.1103/PhysRevB.35.3629}
  {\path{doi:10.1103/PhysRevB.35.3629}}.

\bibitem{Kerr1990}
W.~C. Kerr, P.~S. Lomdahl, Quantum-mechanical derivation of the {D}avydov
  equations for multi-quanta states, in: P.~L. Christiansen, A.~C. Scott
  (Eds.), Davydov's Soliton Revisited: Self-Trapping of Vibrational Energy in
  Protein, Springer, New York, 1990, pp. 23--30.
\newblock \href {https://doi.org/10.1007/978-1-4757-9948-4_2}
  {\path{doi:10.1007/978-1-4757-9948-4_2}}.

\bibitem{Barclay2020}
C.~J. Barclay, D.~S. Loiselle, An equivocal final link -- quantitative
  determination of the thermodynamic efficiency of {ATP} hydrolysis -- sullies
  the chain of electric, ionic, mechanical and metabolic steps underlying
  cardiac contraction, Frontiers in Physiology 11 (2020) 183.
\newblock \href {https://doi.org/10.3389/fphys.2020.00183}
  {\path{doi:10.3389/fphys.2020.00183}}.

\bibitem{Kammermeier1982}
H.~Kammermeier, P.~Schmidt, E.~J\"{u}ngling, Free energy change of
  {ATP}-hydrolysis: a causal factor of early hypoxic failure of the
  myocardium?, Journal of Molecular and Cellular Cardiology 14~(5) (1982)
  267--277.
\newblock \href {https://doi.org/10.1016/0022-2828(82)90205-X}
  {\path{doi:10.1016/0022-2828(82)90205-X}}.

\bibitem{Weiss2005}
R.~G. Weiss, G.~Gerstenblith, P.~A. Bottomley, {ATP} flux through creatine
  kinase in the normal, stressed, and failing human heart, Proceedings of the
  National Academy of Sciences of the United States of America 102~(3) (2005)
  808--813.
\newblock \href {https://doi.org/10.1073/pnas.0408962102}
  {\path{doi:10.1073/pnas.0408962102}}.

\bibitem{Ostrovskaya2001}
E.~A. Ostrovskaya, S.~F. Mingaleev, Y.~S. Kivshar, Y.~B. Gaididei, P.~L.
  Christiansen, Multi-soliton energy transport in anharmonic lattices, Physics
  Letters A 282~(3) (2001) 157--162.
\newblock \href {https://doi.org/10.1016/S0375-9601(01)00157-8}
  {\path{doi:10.1016/S0375-9601(01)00157-8}}.

\bibitem{Chen2017}
Y.~Chen, Z.~Yan, D.~Mihalache, B.~A. Malomed, Families of stable solitons and
  excitations in the {PT}-symmetric nonlinear {S}chr\"{o}dinger equations with
  position-dependent effective masses, Scientific Reports 7~(1) (2017) 1257.
\newblock \href {https://doi.org/10.1038/s41598-017-01401-3}
  {\path{doi:10.1038/s41598-017-01401-3}}.

\bibitem{Vakhnenko2021}
O.~O. Vakhnenko, Nonlinear integrable dynamics of coupled vibrational and
  intra-site excitations on a regular one-dimensional lattice, Physics Letters
  A 405 (2021) 127431.
\newblock \href {https://doi.org/10.1016/j.physleta.2021.127431}
  {\path{doi:10.1016/j.physleta.2021.127431}}.

\bibitem{Georgiev2020b}
D.~D. Georgiev, J.~F. Glazebrook, Launching of {D}avydov solitons in protein
  $\alpha$-helix spines, Physica E: Low-dimensional Systems and Nanostructures
  124 (2020) 114332.
\newblock \href {https://doi.org/10.1016/j.physe.2020.114332}
  {\path{doi:10.1016/j.physe.2020.114332}}.

\bibitem{Behrmann2012}
E.~Behrmann, M.~M\"{u}ller, P.~A. Penczek, H.~G. Mannherz, D.~J. Manstein,
  S.~Raunser, Structure of the rigor actin-tropomyosin-myosin complex, Cell
  150~(2) (2012) 327--338.
\newblock \href {https://doi.org/10.1016/j.cell.2012.05.037}
  {\path{doi:10.1016/j.cell.2012.05.037}}.

\bibitem{Capaldi2002}
R.~A. Capaldi, R.~Aggeler, Mechanism of the {F}$_1${F}$_0$-type {ATP} synthase,
  a biological rotary motor, Trends in Biochemical Sciences 27~(3) (2002)
  154--160.
\newblock \href {https://doi.org/10.1016/S0968-0004(01)02051-5}
  {\path{doi:10.1016/S0968-0004(01)02051-5}}.

\bibitem{Catterall1995}
W.~A. Catterall, Structure and function of voltage-gated ion channels, Annual
  Review of Biochemistry 64~(1) (1995) 493--531.
\newblock \href {https://doi.org/10.1146/annurev.bi.64.070195.002425}
  {\path{doi:10.1146/annurev.bi.64.070195.002425}}.

\bibitem{Liu2012}
H.-L. Liu, C.~W. Pemble~Iv, S.~A. Endow, Neck-motor interactions trigger
  rotation of the kinesin stalk, Scientific Reports 2~(1) (2012) 236.
\newblock \href {https://doi.org/10.1038/srep00236}
  {\path{doi:10.1038/srep00236}}.

\bibitem{Zhou2015}
Q.~Zhou, Y.~Lai, T.~Bacaj, M.~Zhao, A.~Y. Lyubimov, M.~Uervirojnangkoorn, O.~B.
  Zeldin, A.~S. Brewster, N.~K. Sauter, A.~E. Cohen, S.~M. Soltis,
  R.~Alonso-Mori, M.~Chollet, H.~T. Lemke, R.~A. Pfuetzner, U.~B. Choi, W.~I.
  Weis, J.~Diao, T.~C. S\"{u}dhof, A.~T. Brunger, Architecture of the
  synaptotagmin-{SNARE} machinery for neuronal exocytosis, Nature 525~(7567)
  (2015) 62--67.
\newblock \href {https://doi.org/10.1038/nature14975}
  {\path{doi:10.1038/nature14975}}.

\bibitem{Zhang1998}
B.~Zhang, Y.~H. Koh, R.~B. Beckstead, V.~Budnik, B.~Ganetzky, H.~J. Bellen,
  Synaptic vesicle size and number are regulated by a clathrin adaptor protein
  required for endocytosis, Neuron 21~(6) (1998) 1465--1475.
\newblock \href {https://doi.org/10.1016/S0896-6273(00)80664-9}
  {\path{doi:10.1016/S0896-6273(00)80664-9}}.

\bibitem{Jiang2019}
R.~Jiang, S.~Vandal, S.~Park, S.~Majd, E.~T\"{u}zel, W.~O. Hancock, Microtubule
  binding kinetics of membrane-bound kinesin-1 predicts high motor copy numbers
  on intracellular cargo, Proceedings of the National Academy of Sciences
  116~(52) (2019) 26564--26570.
\newblock \href {https://doi.org/10.1073/pnas.1916204116}
  {\path{doi:10.1073/pnas.1916204116}}.

\bibitem{Chanda2008}
B.~Chanda, F.~Bezanilla, A common pathway for charge transport through
  voltage-sensing domains, Neuron 57~(3) (2008) 345--351.
\newblock \href {https://doi.org/10.1016/j.neuron.2008.01.015}
  {\path{doi:10.1016/j.neuron.2008.01.015}}.

\bibitem{Tauber1963}
S.~Tauber, On multinomial coefficients, The American Mathematical Monthly
  70~(10) (1963) 1058--1063.
\newblock \href {https://doi.org/10.1080/00029890.1963.11992172}
  {\path{doi:10.1080/00029890.1963.11992172}}.

\bibitem{Fock1932}
V.~A. Fock, Konfigurationsraum und zweite {Q}uantelung, Zeitschrift f\"{u}r
  Physik 75~(9) (1932) 622--647.
\newblock \href {https://doi.org/10.1007/bf01344458}
  {\path{doi:10.1007/bf01344458}}.

\bibitem{Faddeev2004}
L.~D. Faddeev, L.~A. Khalfin, I.~V. Komarov, V. A. Fock -- Selected Works:
  Quantum Mechanics and Quantum Field Theory, Chapman \& Hall/CRC, Boca Raton,
  2004.
\newblock \href {https://doi.org/10.1201/9780203643204}
  {\path{doi:10.1201/9780203643204}}.

\bibitem{Gross2004}
F.~Gross, Relativistic Quantum Mechanics and Field Theory, Wiley-VCH, Weinheim,
  2004.

\bibitem{Wang2008}
E.~T. Wang, R.~Sandberg, S.~Luo, I.~Khrebtukova, L.~Zhang, C.~Mayr, S.~F.
  Kingsmore, G.~P. Schroth, C.~B. Burge, Alternative isoform regulation in
  human tissue transcriptomes, Nature 456~(7221) (2008) 470--476.
\newblock \href {https://doi.org/10.1038/nature07509}
  {\path{doi:10.1038/nature07509}}.

\bibitem{Chen2009}
M.~Chen, J.~L. Manley, Mechanisms of alternative splicing regulation: insights
  from molecular and genomics approaches, Nature Reviews Molecular Cell Biology
  10~(11) (2009) 741--754.
\newblock \href {https://doi.org/10.1038/nrm2777} {\path{doi:10.1038/nrm2777}}.

\bibitem{Nilsen2010}
T.~W. Nilsen, B.~R. Graveley, Expansion of the eukaryotic proteome by
  alternative splicing, Nature 463~(7280) (2010) 457--463.
\newblock \href {https://doi.org/10.1038/nature08909}
  {\path{doi:10.1038/nature08909}}.

\bibitem{Shi2017}
Y.~Shi, Mechanistic insights into precursor messenger {RNA} splicing by the
  spliceosome, Nature Reviews Molecular Cell Biology 18~(11) (2017) 655--670.
\newblock \href {https://doi.org/10.1038/nrm.2017.86}
  {\path{doi:10.1038/nrm.2017.86}}.

\bibitem{Caremani2015}
M.~Caremani, L.~Melli, M.~Dolfi, V.~Lombardi, M.~Linari, Force and number of
  myosin motors during muscle shortening and the coupling with the release of
  the {ATP} hydrolysis products, Journal of Physiology 593~(15) (2015)
  3313--3332.
\newblock \href {https://doi.org/10.1113/jp270265}
  {\path{doi:10.1113/jp270265}}.

\bibitem{Min2013}
D.~Min, K.~Kim, C.~Hyeon, Y.~Hoon~Cho, Y.-K. Shin, T.-Y. Yoon, Mechanical
  unzipping and rezipping of a single snare complex reveals hysteresis as a
  force-generating mechanism, Nature Communications 4~(1) (2013) 1705.
\newblock \href {https://doi.org/10.1038/ncomms2692}
  {\path{doi:10.1038/ncomms2692}}.

\bibitem{Shohet1989}
R.~V. Shohet, M.~A. Conti, S.~Kawamoto, Y.~A. Preston, D.~A. Brill, R.~S.
  Adelstein, Cloning of the {cDNA} encoding the myosin heavy chain of a
  vertebrate cellular myosin, Proceedings of the National Academy of Sciences
  86~(20) (1989) 7726--7730.
\newblock \href {https://doi.org/10.1073/pnas.86.20.7726}
  {\path{doi:10.1073/pnas.86.20.7726}}.

\bibitem{Kivshar1987}
Y.~S. Kivshar, A.~M. Kosevich, O.~A. Chubykalo, Resonant and non-resonant
  soliton scattering by impurities, Physics Letters A 125~(1) (1987) 35--40.
\newblock \href {https://doi.org/10.1016/0375-9601(87)90514-7}
  {\path{doi:10.1016/0375-9601(87)90514-7}}.

\bibitem{Georgiev2022b}
D.~D. Georgiev, S.~P. Gudder, Sensitivity of entanglement measures in bipartite
  pure quantum states, Modern Physics Letters B (2022) 2250101\href
  {https://doi.org/10.1142/S0217984922501019}
  {\path{doi:10.1142/S0217984922501019}}.

\bibitem{Davydov1986b}
A.~S. Davydov, Solitons in biology, in: S.~E. Trullinger, V.~E. Zakharov, V.~L.
  Pokrovsky (Eds.), Solitons, Modern Problems in Condensed Matter Sciences,
  North-Holland, Amsterdam, 1986, pp. 1--51.
\newblock \href {https://doi.org/10.1016/B978-0-444-87002-5.50007-2}
  {\path{doi:10.1016/B978-0-444-87002-5.50007-2}}.

\bibitem{Walker2009}
J.~M. Walker, The Protein Protocols Handbook, 3rd Edition, Humana Press,
  Totowa, New Jersey, 2009.
\newblock \href {https://doi.org/10.1007/978-1-59745-198-7}
  {\path{doi:10.1007/978-1-59745-198-7}}.

\bibitem{Staudt2011}
A.-C. Staudt, S.~Wenkel, Regulation of protein function by `micro{P}roteins',
  EMBO Reports 12~(1) (2011) 35--42.
\newblock \href {https://doi.org/10.1038/embor.2010.196}
  {\path{doi:10.1038/embor.2010.196}}.

\bibitem{Hilton2012}
G.~R. Hilton, J.~L.~P. Benesch, Two decades of studying non-covalent
  biomolecular assemblies by means of electrospray ionization mass
  spectrometry, Journal of The Royal Society Interface 9~(70) (2012) 801--816.
\newblock \href {https://doi.org/10.1098/rsif.2011.0823}
  {\path{doi:10.1098/rsif.2011.0823}}.

\bibitem{Chorev2015}
D.~S. Chorev, G.~Ben-Nissan, M.~Sharon, Exposing the subunit diversity and
  modularity of protein complexes by structural mass spectrometry approaches,
  Proteomics 15~(16) (2015) 2777--2791.
\newblock \href {https://doi.org/10.1002/pmic.201400517}
  {\path{doi:10.1002/pmic.201400517}}.

\bibitem{Shon2018}
M.~J. Shon, H.~Kim, T.-Y. Yoon, Focused clamping of a single neuronal {SNARE}
  complex by complexin under high mechanical tension, Nature Communications
  9~(1) (2018) 3639.
\newblock \href {https://doi.org/10.1038/s41467-018-06122-3}
  {\path{doi:10.1038/s41467-018-06122-3}}.

\bibitem{Georgiev2018}
D.~D. Georgiev, J.~F. Glazebrook, The quantum physics of synaptic communication
  via the {SNARE} protein complex, Progress in Biophysics and Molecular Biology
  135 (2018) 16--29.
\newblock \href {https://doi.org/10.1016/j.pbiomolbio.2018.01.006}
  {\path{doi:10.1016/j.pbiomolbio.2018.01.006}}.

\bibitem{Guo2022}
H.~Guo, J.~L. Rubinstein, Structure of {ATP} synthase under strain during
  catalysis, Nature Communications 13~(1) (2022) 2232.
\newblock \href {https://doi.org/10.1038/s41467-022-29893-2}
  {\path{doi:10.1038/s41467-022-29893-2}}.

\bibitem{Weber2000}
J.~Weber, A.~E. Senior, {ATP} synthase: what we know about {ATP} hydrolysis and
  what we do not know about {ATP} synthesis, Biochimica et Biophysica Acta
  (BBA) - Bioenergetics 1458~(2) (2000) 300--309.
\newblock \href {https://doi.org/10.1016/S0005-2728(00)00082-7}
  {\path{doi:10.1016/S0005-2728(00)00082-7}}.

\bibitem{Devi2015}
S.~K. Devi, V.~P.~R. Chichili, J.~Jeyakanthan, D.~Velmurugan, J.~Sivaraman,
  Structural basis for the hydrolysis of {ATP} by a nucleotide binding subunit
  of an amino acid {ABC} transporter from {T}hermus thermophilus, Journal of
  Structural Biology 190~(3) (2015) 367--372.
\newblock \href {https://doi.org/10.1016/j.jsb.2015.04.012}
  {\path{doi:10.1016/j.jsb.2015.04.012}}.

\bibitem{Vorobiev2003}
S.~Vorobiev, B.~Strokopytov, D.~G. Drubin, C.~Frieden, S.~Ono, J.~Condeelis,
  P.~A. Rubenstein, S.~C. Almo, The structure of nonvertebrate actin:
  Implications for the {ATP} hydrolytic mechanism, Proceedings of the National
  Academy of Sciences 100~(10) (2003) 5760--5765.
\newblock \href {https://doi.org/10.1073/pnas.0832273100}
  {\path{doi:10.1073/pnas.0832273100}}.

\bibitem{Cross2016}
R.~A. Cross, Mechanochemistry of the kinesin-1 {ATP}ase, Biopolymers 105~(8)
  (2016) 476--482.
\newblock \href {https://doi.org/10.1002/bip.22862}
  {\path{doi:10.1002/bip.22862}}.

\end{thebibliography}
\end{document}